\documentclass[a4paper,11pt]{article}
\pdfoutput=1 

\usepackage{jheppub} 

\usepackage[T1]{fontenc} 
\usepackage{slashed}
\usepackage{tikz}
\usetikzlibrary{shapes,arrows,positioning,automata,backgrounds,calc,er,patterns}
\usepackage{tikz-feynman}
\tikzfeynmanset{compat=1.1.0}
\usepackage{xcolor}
\usepackage{makecell}
\usepackage{rotating}
\usepackage{pdflscape}
\usepackage{float}
\usepackage{multicol}
\usepackage{xspace}
\usepackage{booktabs}
\usepackage{xspace}
\usepackage{pifont}
\usepackage[shortlabels]{enumitem}
\usepackage{amssymb, latexsym, amsmath, amsthm}

\DeclareMathAlphabet\mathbfcal{OMS}{cmsy}{b}{n}

\usepackage{tikz}
\usetikzlibrary{shapes.geometric, arrows, positioning}
\tikzstyle{fullspace} = [rectangle, minimum width=4cm, minimum height=1.5cm, text centered, text width=3cm, draw=black, fill=white]
\tikzstyle{subspace} = [rectangle, rounded corners, minimum width=4cm, minimum height=1.5cm,text centered, text width=3cm, draw=black, fill=white]
\tikzstyle{arrow} = [thick,->,>=stealth]
\tikzstyle{circlebox} = [circle,text centered, minimum width = 1.5 cm, minimum height = 1.5 cm, draw=black,fill=white]
\tikzstyle{recbox} = [rectangle,text centered, minimum width = 1.5 cm, minimum height = 1.5 cm, draw=black,fill=white]

\tikzstyle{startstop} = [rectangle, rounded corners, 
minimum width=4cm, 
minimum height=1cm,
text centered, 
draw=black, 
fill=red!30]

\tikzstyle{io} = [trapezium, 
trapezium stretches=true, 
trapezium left angle=70, 
trapezium right angle=110, 
minimum width=1.5cm, 
minimum height=1cm, text centered, 
draw=black, fill=blue!30]

\tikzstyle{process} = [rectangle, 
minimum width=1cm, 
minimum height=1cm, 
text centered, 
draw=black, 
fill=orange!30]

\tikzstyle{decision} = [diamond, 
minimum width=3cm, 
minimum height=1cm, 
text centered, 
draw=black, 
fill=green!30]

\newcommand{\ome}{(1-\epsilon)}
\newcommand{\e}{\epsilon}

\newcommand{\modsq}[1]{\ensuremath{\left\vert #1 \right\vert^2}}
\newcommand{\calM}{\ensuremath{\mathcal{M}}}
\newcommand{\order}[1]{\ensuremath{\mathcal{O}}\left( #1 \right)\xspace}

\newcommand{\NthreeLO}{N$^3$LO\xspace}

\newcommand{\Pqgzero}{P_{qg}^{(0)}}

\newcommand{\Pggzero}{P_{gg}^{(0)}}
\newcommand{\Pqqzero}{P_{q\bar{q}}^{(0)}}
\newcommand{\PggSzero}{P_{gg}^{\text{sub},(0)}}

\newcommand{\Pqg}{P_{qg}^{(0)}}

\newcommand{\Pgg}{P_{gg}^{(0)}}
\newcommand{\Pqq}{P_{q\bar{q}}^{(0)}}

\newcommand{\PggS}{P_{gg}^{\text{sub},(0)}}

\newcommand{\Bz}{\beta_0}
\newcommand{\bz}{b_0}
\newcommand{\bzf}{b_{0,F}}

\newcommand{\Sgone}{S_{g}^{(1)}}
\newcommand{\Sgtone}{\widetilde{S}_{g}^{(1)}}
\newcommand{\Sghone}{\widehat{S}_{g}^{(1)}}
\newcommand{\Sgzero}{S_{g}^{(0)}}
\newcommand{\Pqgone}{P_{qg}^{(1)}}
\newcommand{\Pqgtone}{\widetilde{P}_{qg}^{(1)}}
\newcommand{\Pqghone}{\widehat{P}_{qg}^{(1)}}

\newcommand{\Pggone}{P_{gg}^{(1)}}
\newcommand{\Pggtone}{\widetilde{P}_{gg}^{(1)}}
\newcommand{\Pgghone}{\widehat{P}_{gg}^{(1)}}
\newcommand{\Pggonesub}{P_{gg}^{(1),\mathrm{sub}}}

\newcommand{\Pgghonesub}{\widehat{P}_{gg}^{(1),\mathrm{sub}}}
\newcommand{\Pqqone}{P_{q\bar{q}}^{(1)}}
\newcommand{\Pqqtone}{\widetilde{P}_{q\bar{q}}^{(1)}}
\newcommand{\Pqqhone}{\widehat{P}_{q\bar{q}}^{(1)}}

\newcommand{\PSup}{\mathrm{\mathbf{S}}^\uparrow\xspace}
\newcommand{\PCup}{\mathrm{\mathbf{C}}^\uparrow\xspace}
\newcommand{\PXup}{\mathrm{\mathbf{P}}_X^\uparrow\xspace}
\newcommand{\PXdown}{\mathrm{\mathbf{P}}_X^\downarrow\xspace}

\newcommand{\PSdown}{\mathrm{\mathbf{S}}^\downarrow\xspace}
\newcommand{\PCdown}{\mathrm{\mathbf{C}}^\downarrow\xspace}

\newcommand{\PPup}{\mathrm{\mathbf{P}}^\uparrow\xspace}
\newcommand{\PPdown}{\mathrm{\mathbf{P}}^\downarrow\xspace}
\newcommand{\Peup}{\mathrm{\mathbf{P}}_\e^\uparrow\xspace}
\newcommand{\Pedown}{\mathrm{\mathbf{P}}_\e^\downarrow\xspace}

\newcommand{\XLO}{X_3^0}
\newcommand{\X}{X_4^0}

\newcommand{\A}{A_4^0}

\newcommand{\D}{D_4^0}

\newcommand{\Arv}{A_3^1}
\newcommand{\Atrv}{\widetilde{A}_3^1}
\newcommand{\Ahrv}{\widehat{A}_3^1}
\newcommand{\Drv}{D_3^1}

\newcommand{\Dtrv}{\widetilde{D}_3^1}
\newcommand{\Dhrv}{\widehat{D}_3^1}
\newcommand{\Erv}{E_3^1}
\newcommand{\Etrv}{\widetilde{E}_3^1}
\newcommand{\Ehrv}{\widehat{E}_3^1}
\newcommand{\Frv}{F_3^1}
\newcommand{\Fhrv}{\widehat{F}_3^1}
\newcommand{\Grv}{G_3^1}
\newcommand{\Gtrv}{\widetilde{G}_3^1}
\newcommand{\Ghrv}{\widehat{G}_3^1}

\newcommand{\Arvold}{A_3^{1,\oldant}}
\newcommand{\Atrvold}{\widetilde{A}_3^{1,\oldant}}
\newcommand{\Ahrvold}{\widehat{A}_1^{1,\oldant}}
\newcommand{\Drvold}{D_3^{1,\oldant}}
\newcommand{\Dtrvold}{\widetilde{D}_3^{1,\oldant}}
\newcommand{\Dhrvold}{\widehat{D}_3^{1,\oldant}}
\newcommand{\Ehrvold}{\widehat{E}_3^{1,\oldant}}
\newcommand{\Ervold}{E_3^{1,\oldant}}
\newcommand{\Etrvold}{\widetilde{E}_3^{1,\oldant}}
\newcommand{\Frvold}{F_3^{1,\oldant}}
\newcommand{\Gtrvold}{\widetilde{G}_3^{1,\oldant}}
\newcommand{\Fhrvold}{\widehat{F}_3^{1,\oldant}}
\newcommand{\Grvold}{G_3^{1,\oldant}}
\newcommand{\Ghrvold}{\widehat{G}_3^{1,\oldant}}

\newcommand{\calArvold}{{\cal A}_3^{1,\oldant}}
\newcommand{\calAtrvold}{\widetilde{\cal A}_3^{1,\oldant}}
\newcommand{\calAhrvold}{\widehat{\cal A}_1^{1,\oldant}}
\newcommand{\calDrvold}{{\cal D}_3^{1,\oldant}}
\newcommand{\calDhrvold}{\widehat{\cal D}_3^{1,\oldant}}
\newcommand{\calEhrvold}{\widehat{\cal E}_3^{1,\oldant}}
\newcommand{\calErvold}{{\cal E}_3^{1,\oldant}}
\newcommand{\calEtrvold}{\widetilde{\cal E}_3^{1,\oldant}}
\newcommand{\calFrvold}{{\cal F}_3^{1,\oldant}}

\newcommand{\calFhrvold}{\widehat{\cal F}_3^{1,\oldant}}

\newcommand{\calArv}{{\cal A}_3^1}
\newcommand{\calAtrv}{\widetilde{{\cal A}}_3^1}
\newcommand{\calAhrv}{\widehat{{\cal A}}_3^1}
\newcommand{\calDrv}{{\cal D}_3^1}

\newcommand{\calDtrv}{\widetilde{{\cal D}}_3^1}
\newcommand{\calDhrv}{\widehat{{\cal D}}_3^1}
\newcommand{\calErv}{{\cal E}_3^1}
\newcommand{\calEtrv}{\widetilde{{\cal E}}_3^1}
\newcommand{\calEhrv}{\widehat{{\cal E}}_3^1}
\newcommand{\calFrv}{{\cal F}_3^1}
\newcommand{\calFhrv}{\widehat{{\cal F}}_3^1}
\newcommand{\calGrv}{{\cal G}_3^1}
\newcommand{\calGtrv}{\widetilde{{\cal G}}_3^1}
\newcommand{\calGhrv}{\widehat{{\cal G}}_3^1}

\newcommand{\Dt}{\widetilde{D}_4^0}

\newcommand{\oldant}{\, \mathrm{OLD}}
\newcommand{\Xold}[1]{#1_3^{0,\oldant}}

\newcommand{\Dold}{D_4^{0,\oldant}}

\newcommand{\calA}{{\cal A}_4^0}

\newcommand{\calB}{{\cal B}_4^0}

\newcommand{\calC}{{\cal C}_4^0}

\newcommand{\calD}{{\cal D}_4^0}

\newcommand{\calF}{{\cal F}_4^0}

\newcommand{\calEa}{{\cal E}_4^{0}}
\newcommand{\calEb}{\overline{{\cal E}}_4^{0}}

\newcommand{\calGa}{{\cal G}_4^{0}}
\newcommand{\calGb}{\overline{{\cal G}}_4^{0}}
\newcommand{\calH}{{\cal H}_4^0}

\newcommand{\calAt}{\widetilde{{\cal A}}_4^0}
\newcommand{\calDt}{\widetilde{{\cal D}}_4^0}
\newcommand{\calFt}{\widetilde{{\cal F}}_4^0}
\newcommand{\calEt}{\widetilde{{\cal E}}_4^0}
\newcommand{\calGt}{\widetilde{{\cal G}}_4^0}

\newcommand{\cala}{{\cal A}_3^0}
\newcommand{\cald}{{\cal D}_3^0}
\newcommand{\calf}{{\cal F}_3^0}
\newcommand{\cale}{{\cal E}_3^0}
\newcommand{\calg}{{\cal G}_3^0}

\newcommand{\Ssoft}{\mathrm{Ssoft}}
\newcommand{\Scol}{\mathrm{Scol}}

\newcommand{\Scolt}{\widetilde{\mathrm{Scol}}}

\newcommand{\Scolh}{\widehat{\mathrm{Scol}}}

\newcommand{\FinPoles}{\text{TPoles}}
\newcommand{\FinPolest}{\widetilde{\text{TPoles}}}
\newcommand{\FinPolesh}{\widehat{\text{TPoles}}}
\newcommand{\Poles}{{\mathcal Poles}}

\newcommand{\calSsoft}{\mathcal{S}\kern-0.05em{s\kern-0.05em o\kern-0.05em f\kern-0.05em t}}
\newcommand{\calScol}{\mathcal{S}\kern-0.05em{c\kern-0.05em o\kern-0.05em l}}
\newcommand{\calDsoft}{\mathcal{D}\kern-0.05em{s\kern-0.05em o\kern-0.05em f\kern-0.05em t}}
\newcommand{\calTcol}{\mathcal{T}\kern-0.05em{c\kern-0.05em o\kern-0.05em l}}
\newcommand{\calDcol}{\mathcal{D}\kern-0.05em{c\kern-0.05em o\kern-0.05em l}}

\newcommand{\qb}{\bar{q}}

\newcommand{\omz}{(1-z)}
\newcommand{\Li}{\mathrm{Li}}

\newcommand{\Af}{R_{\e}}

\newcommand{\omxj}{(1-x_j)}

\renewcommand{\xi}{x_{i}}
\newcommand{\xj}{x_{j}}

\newcommand{\Jcol}[1]{\mathbfcal{J}^{(#1)}}
\newcommand{\Jcolb}[1]{\overline{\mathbfcal{J}}^{(#1)}}

\newcommand{\Jtot}[1]{\mathcal{J}^{(#1)}_2}
\newcommand{\Jtotb}[1]{\overline{\mathcal{J}}^{(#1)}_2}
\newcommand{\JtotOLD}[1]{\mathcal{J}^{(#1),\oldant}_2}
\newcommand{\JtotbOLD}[1]{\overline{\mathcal{J}}^{(#1),\oldant}_2}

\newcommand{\J}[1]{J_{2}^{(#1)}}
\newcommand{\Jh}[1]{\widehat{J_{2}}^{(#1)}}
\newcommand{\Jt}[1]{\widetilde{J_{2}}^{(#1)}}
\newcommand{\Jht}[1]{\widehat{\widetilde{J_{2}}}^{(#1)}}
\newcommand{\Jhh}[1]{\widehat{\widehat{J_{2}}}^{(#1)}}
\newcommand{\Jb}[1]{\overline{J}_{2}^{(#1)}}

\title{
A general algorithm to build mixed real and virtual antenna
functions for higher-order calculations 
}

\author[1,2]{Oscar Braun-White,}
\author[1,2]{Nigel Glover,}
\author[3]{Christian T Preuss}

\affiliation[1]{Institute for Particle Physics Phenomenology, Department of Physics, Durham University, South Road, Durham, DH1 3LE, UK}
\affiliation[2]{Physik-Institut, Universität Zurich, Winterthurerstrasse 190, CH-8057 Zürich, Switzerland}
\affiliation[3]{Institute for Theoretical Physics, ETH, CH-8093 Zürich, Wolfgang-Pauli-Strasse 27, Switzerland}

\emailAdd{oscar.r.braun-white@durham.ac.uk}
\emailAdd{e.w.n.glover@durham.ac.uk}
\emailAdd{cpreuss@phys.ethz.ch}

\preprint{IPPP/23/31, ZU-TH 30/23}

\abstract{
The antenna-subtraction technique has demonstrated remarkable effectiveness in providing next-to-next-to-leading order in $\alpha_s$ (NNLO) predictions for a wide range of processes relevant for the Large Hadron Collider.  In a previous paper \cite{paper2}, we demonstrated how to build real-radiation antenna functions for any number of real emissions directly from a specified list of unresolved limits. Here, we extend this procedure to the mixed case of real and virtual radiation, for any number of real and virtual emissions.  A novel feature of the algorithm is the requirement to match the antenna constructed with the correct unresolved limits to the other elements of the subtraction scheme.  We discuss how this can be achieved and provide a full set of real-virtual NNLO antenna functions (together with their integration over the final-final unresolved phase space).  We demonstrate that these antennae can be combined with the real-radiation antennae of Ref.~\cite{paper2} to form a consistent NNLO subtraction scheme that cancels all explicit and implicit singularities at NNLO. We anticipate that the improved antenna functions should be more amenable to automation, thereby making the construction of subtraction terms for more complicated processes simpler at NNLO.
}

\begin{document}
\maketitle
\flushbottom

\section{Introduction}

The Large Hadron Collider (LHC) offers an unprecedented opportunity to scrutinise a wide range of observables involving Higgs bosons, electroweak bosons, top quarks, and hadronic jets with remarkable accuracy. Through precise experimental measurements, we can directly investigate the fundamental interactions of elementary particles at short distances, pushing the boundaries of our knowledge and providing valuable insights into the fundamental interactions that govern the universe.  The exploration of LHC physics, particularly in the absence of new particle discoveries, holds immense significance. By scrutinising the LHC data with high precision, even the slightest deviations from the predictions of the Standard Model (SM) can have profound implications for our understanding of the natural world. Such small deviations in measurements have the potential to revolutionise our knowledge and guide us towards physics beyond the Standard Model. Hence, precision phenomenology emerges as a crucial component in the quest for new physics.

With the anticipated dataset from the High-Luminosity LHC, the statistical uncertainties on many observables will be negligible and percent-level accuracy is likely to be achieved experimentally.  Achieving similar percent-level accuracy for theoretical predictions requires advancements in fixed-order calculations, parton distribution functions, parton showers, and the modeling of non-perturbative effects. Ongoing progress is being made in all these areas. In the realm of perturbative Quantum Chromodynamics (QCD), reaching the desired level of refinement typically involves extending fixed-order calculations to at least next-to-next-to-leading order (NNLO) in the strong-coupling expansion.

However, higher-order calculations demand special attention due to the intricate interplay between real and virtual corrections across different-multiplicity phase spaces~\cite{Kinoshita,LeeNauenberg}. Implicit infrared divergences arise from unresolved real radiation, such as soft or collinear emissions, and are ultimately cancelled by explicit poles in the virtual matrix elements.  This cancellation takes place through integration over the relevant unresolved phase space. Subtraction schemes are currently regarded as the most elegant solution to address these complexities.

At the next-to-leading order (NLO) level, schemes like Catani-Seymour dipole subtraction~\cite{Catani:1996vz} and FKS subtraction~\cite{Frixione:1995ms} were developed in the mid-1990's. Together with automated one-loop matrix-element generators~\cite{madgraph:2011uj,Cascioli:2011va}, these schemes are used for fully-differential high-multiplicity processes. NLO matching schemes such as MC@NLO~\cite{Frixione:2002ik} and POWHEG~\cite{Nason:2004rx,Frixione:2007vw} have been developed which systematically combine NLO fixed-order calculations with all-order parton-shower resummation. These innovations laid the foundations for the state-of-the-art multi-purpose event generators~\cite{powheg:2010xd,madgraph:2011uj,Bellm:2019zci,Sherpa:2019gpd,Bierlich:2022pfr}, see Ref.~\cite{Campbell:2022qmc} for a review. 

At NNLO, the situation is less advanced. Despite recent progress, two-loop matrix elements represent significant challenges often requiring bespoke integral reduction relations and the evaluation of new master integrals.  At the same time, the pattern of cancellation of infrared divergences across the different-multiplicity final states is much more complicated.  Several subtraction schemes have been devised~\cite{Gehrmann-DeRidder:2005btv,Boughezal:2011jf,DelDuca:2016ily,Caola:2017dug,Magnea:2018hab,Herzog:2018ily,TorresBobadilla:2020ekr} and the implementation of these methods is currently done one process at a time.   They do not straightforwardly scale to higher multiplicities. 

Among the various methods used for fully-differential NNLO calculations in QCD, the antenna-subtraction scheme~\cite{Gehrmann-DeRidder:2005btv,Gehrmann-DeRidder:2005alt,Gehrmann-DeRidder:2005svg} has proven to be highly successful. Initially proposed for electron-positron annihilation with massless partons, it enabled the calculation of NNLO corrections to 3-jet production and related event-shape observables at LEP energies~\cite{Catani:2007vq,Gehrmann-DeRidder:2007foh,Gehrmann-DeRidder:2007nzq,Gehrmann-DeRidder:2007vsv,Gehrmann-DeRidder:2008qsl}. The scheme was extended to handle initial-state radiation relevant to processes with initial-state hadrons~\cite{Daleo:2006xa,Daleo:2009yj,Pires:2010jv,Boughezal:2010mc,Gehrmann:2011wi,Gehrmann-DeRidder:2012too,Currie:2013vh} and has now been applied to a range of LHC processes through the parton-level NNLOJET code. The extension of antenna subtraction for the production of heavy particles at hadron colliders has been studied in Refs. \cite{Gehrmann-DeRidder:2009lyc,Abelof:2011ap,Bernreuther:2011jt,Abelof:2011jv,Abelof:2012bga,Abelof:2012rv,Bernreuther:2013uma,Dekkers:2014hna}. 
Besides its application in fixed-order calculations, the antenna framework has also been utilised in antenna-shower algorithms \cite{Gustafson:1987rq,Lonnblad:1992tz,Giele:2007di,Giele:2011cb,Fischer:2016vfv,Brooks:2020upa}, where it enabled proof-of-concept frameworks for higher-order corrections \cite{Li:2016yez} and fully-differential NNLO matching \cite{Campbell:2021svd}.

At \NthreeLO, inclusive~\cite{Anastasiou:2015vya,Anastasiou:2016cez,Mistlberger:2018etf,Dreyer:2016oyx,Duhr:2019kwi,Duhr:2020kzd,Chen:2019lzz,Currie:2018fgr,Dreyer:2018qbw,Duhr:2020sdp,Duhr:2020seh} as well as more differential calculations have started to emerge~\cite{Dulat:2017prg,Dulat:2018bfe,Cieri:2018oms,Chen:2021isd,Chen:2021vtu,Billis:2021ecs,Chen:2022cgv,Neumann:2022lft,Camarda:2021ict,Chen:2022lwc,Baglio:2022wzu}, the latter mainly for $2 \to 1$ processes via the use of the Projection-to-Born method~\cite{Cacciari:2015jma} or $k_\mathrm{T}$-slicing techniques~\cite{Catani:2007vq} to promote established NNLO calculations to \NthreeLO. We note that the first steps towards an \NthreeLO antenna subtraction scheme have been taken in Refs.~\cite{Jakubcik:2022zdi,Chen:2023fba}. Nevertheless, at the moment, calculations for higher multiplicities are currently hindered by the lack of process-independent \NthreeLO subtraction schemes.

In the antenna subtraction scheme, antenna functions are used to subtract specific sets of unresolved singularities, so that a typical subtraction term has the form
\begin{equation} \label{eqn:subterm}
    X_{n+2}^\ell(i_1^h,i_3,\ldots,i_{n+2},i_2^h) 
    \modsq{\calM(\ldots,I_1^h,I_2^h,\ldots)} \, ,
\end{equation}
where $X_{n+2}^\ell$ represents an $\ell$-loop, $(n+2)$-particle antenna,
$i_1^h$ and $i_2^h$ represent the hard radiators, and $i_3$ to $i_{n+2}$ denote the $n$ unresolved particles.   
As the hard radiators may either be in the initial or in the final state, final-final (FF), initial-final (IF), and initial-initial (II) configurations need to be considered in general. $\calM$ is the reduced matrix element, with $n$ fewer particles and where $I_1^h$ and $I_2^h$ represent the particles obtained through an appropriate mapping,
\begin{align}
\{ p_{i_1},p_{i_3},\ldots,p_{i_2} \} \mapsto \{ p_{I_1}, p_{I_2} \}  
\end{align}
with $p_i^{\mu}$ representing the four-momentum of particle $i$. At NLO antennae have $n=1$ and $\ell=0$, at NNLO one needs antennae with $n=2,~\ell = 0$ and with $n=1,~\ell=1$, while at \NthreeLO, one needs antennae with $n=3,~\ell = 0$, with $n=2,~\ell = 1$ and with $n=1,~\ell = 2$. 

In the original formulation of the antenna scheme, the antennae were based on matrix elements describing radiation from processes with two coloured particles: $\gamma^* \to q\bar{q}$, $\tilde{\chi} \to \tilde{g}g$ and $H \to gg$, covering the cases where the coloured particles are massless quarks and gluons.  The corresponding $X_4^0$, $X_3^1$ and $X_3^0$ antennae are therefore perfect subtraction terms for the NNLO contributions to processes with two coloured particles.  It was straightforward to utilise these matrix-element-based antennae for processes with three coloured particles, such as $e^+e^- \to 3$~jets, $pp \to V$+jet, $pp \to H$+jet, and for the leading colour contributions to four coloured particle processes like $pp \to 2$~jets. Pushing to the next step, the full colour $pp \to 2$~jets required significant additional work~\cite{Chen:2022clm}. Going beyond the current state of the art with the matrix-element-based antenna approach is a formidable task. This is because the complexity associated with the subtraction terms becomes increasingly challenging as the particle multiplicity grows.  This complexity stems from two primary reasons.

Firstly, the double-real-radiation antenna functions obtained from matrix elements do not always indicate which particles act as the hard radiators. This is particularly the case for antennae involving gluons. To address this issue, sub-antenna functions are introduced. However, constructing these sub-antenna functions at NNLO is an arduous task and often involves introducing unphysical denominators that complicate the analytic integration of the subtraction term. Additionally, analytic integrals are usually known only for the complete antenna functions. As a result, the assembly of antenna-subtraction terms requires careful manipulation to ensure that the sub-antenna functions combine appropriately to form the full antenna functions before integration.

Secondly, NNLO antenna functions can exhibit spurious limits that need to be eliminated through explicit counter terms. However, these counter terms can introduce further spurious limits themselves. Consequently, this can initiate a complex chain of interdependent subtraction terms that do not necessarily reflect the actual singularity structure of the underlying process.

Both of these issues are obstacles to a full automation of the antenna-subtraction scheme at NNLO~\cite{Chen:2022ktf}. In a recent paper~\cite{paper2}, we addressed these issues. We introduced a general algorithm for building antenna functions directly from a specified set of desired infrared limits with a uniform template, in a way that simplifies the construction of subtraction terms in general, while still being straightforwardly analytically integrable. We then constructed a general algorithm to construct real-radiation antenna functions following strictly these design principles and applied it to the case of single-real and double-real radiation, required for NLO and NNLO calculations. The technique makes use of an iterative procedure to remove overlaps between different singular factors that are subsequently projected into the full phase space. As the technique produces only denominators that match physical propagators, all antenna functions could straightforwardly be integrated analytically, which is a cornerstone of the antenna-subtraction method. 

In this paper, we extend the general algorithm of Ref.~\cite{paper2} to the construction of antennae with $\ell \neq 0$. Unlike in the solely real-radiation case, the mixed real and virtual antenna functions contain both explicit and implicit singularities. To illustrate the algorithm, we construct the real-virtual antennae $(n=1,~\ell=1)$ explicitly. The real-virtual antenna functions are built directly from the relevant one-loop limits, properly accounting for the overlap between different limits. The universal factorisation properties of multi-particle loop amplitudes, when one or more particles are unresolved, have been well studied in the literature~\cite{Bern:1994zx,Bern:1998sc,Kosower:1999rx,Bern:1999ry} and serve as an input to the algorithm. 

In addition to building a full set of new $X_3^1$ antennae with both hard radiators in the final state, we demonstrate that the new antenna functions (along with the $X_3^0$ and $\X$ of Ref.~\cite{paper2}) form a complete NNLO subtraction scheme in which the subtraction terms cancel the explicit singularities in the one- and two-loop matrix elements, without leftover infrared singularities hiding in the matrix elements (either by undercounting or overcounting). This means that the new antenna functions have to satisfy particular constraints. First, the cancellation of poles at the real-virtual level (RV) means that the explicit poles in the $X_3^1$ antenna have to cancel against other RV subtraction terms.  In the antenna scheme, these explicit poles are proportional to $X_3^0$ antennae. Therefore, the $X_3^1$ must have a particular pole structure multiplying an $X_3^0$ antenna function. At the double-virtual level (VV), the combinations of integrated antennae coming from the double-real (RR) and RV levels must match the explicit pole structure of the two-loop matrix elements.  In the antenna scheme, this is encoded through a combination of the $\J{2}$ and $\J{1}$ operators in colour space~\cite{Currie:2013vh}. Provided that the pole structure from the relevant combination of $\J{2}$ and $\J{1}$ is unchanged, the subtraction terms will cancel the explicit poles in the two-loop matrix elements.  

The current approach to automation of antenna subtraction~\cite{Chen:2022ktf} involves a reformulation of the colour-ordered antenna subtraction technique in colour space. This method, known as `colourful antenna subtraction', offers a systematic way to construct antenna subtraction terms by working upwards from the most virtual layer, rather than starting from the maximally real layer and working down. By translating infrared poles of virtual corrections captured by $\J{2}$ and $\J{1}$ into real-radiation dipole insertions in colour space, the method efficiently constitutes subtraction terms for single-real radiation up to one-loop level and for double-real radiation at the tree level. One of the key advantages of this approach is the avoidance of directly handling the divergent behavior of real-emission corrections. This feature represents a significant simplification at NNLO. The double-real subtraction term can be obtained as the final step of a fully automatable procedure, eliminating the need to deal with the involved infrared structure of double-real radiation matrix elements. The completion of a consistent set of improved antenna functions for (double-)real and real-virtual radiation presented here will further reduce the complexity of the subtraction terms, because they avoid the need to subtract spurious limits, and therefore reduce the computational overhead associated with precision calculations.

The paper is structured as follows. We outline the design principles for constructing general $X_{n+2}^\ell$ antenna functions in Section~\ref{sec:principles} as well as the principles for matching to the other elements of an antenna-subtraction scheme. 
We describe the general construction algorithm in Section~\ref{sec:algorithm} and give the specific details for the construction of final-final $X_3^1$ in Section~\ref{sec:X31construction}. The full set of one-loop unresolved limits and target poles for the $X_3^1$ is given in Section~\ref{sec:limits}. Using the previous sections, we illustrate the algorithm by explicitly constructing a full set of $X_3^1$ real-virtual antenna functions for hard radiators in the final state in Section~\ref{sec:X31}. Finally, we define the $\J{2}$ and $\J{1}$ operators in this NNLO antenna-subtraction scheme (out of the new $\{X_3^0,\X,X_3^1\}$) and compare their pole structure to the generic VV pole structures in Section~\ref{sec:J22}. This demonstrates that the new subtraction terms will cancel the explicit poles in the two-loop matrix elements and form a complete NNLO subtraction scheme. We conclude and give an outlook on further work in Section~\ref{sec:outlook}. For the sake of completeness, we also enclose appendices listing the tree-level single-unresolved limits, details of the analytic integration over the final-state antenna phase space, and a list of integrated $X_3^1$ antenna that are based on the $X_3^0$ antenna of Ref.~\cite{Gehrmann-DeRidder:2005svg}.

\section{Design principles}
\label{sec:principles}

Within the antenna-subtraction framework, subtraction terms are constructed using antenna functions that describe the unresolved partonic radiation (both soft and collinear) emitted from a pair of hard radiator partons. The construction of an antenna-subtraction term typically involves the following elements:
\begin{itemize}
\item antennae composed of two hard radiators that accurately capture the infrared singularities arising from the emission of $n$ unresolved partons;
\item an on-shell momentum mapping that ensures that the invariant mass of the antenna is preserved while producing the on-shell momenta that appear in the ``reduced'' matrix element; and
\item a	colour factor associated with the specific process and antenna.
\end{itemize}
The latter two items on this list have been solved for general processes, while the first is subject of the current paper and our previous paper~\cite{paper2}.

In the following, we will describe the design principles we impose upon a general idealised $X_{n+2}^\ell$ antenna function, with at least one loop. As opposed to the $\ell=0$ case, antenna functions with additional virtual elements contain explicit poles in the dimensional-regularisation parameter, $\e$. We therefore impose two different sets of design principles: the generic design principles discussed in Section~\ref{sec:generalprinciples}; and the antenna-scheme-dependent design principles discussed in Section~\ref{sec:matching}.
The former principles ensure that the antenna function has the correct infrared limits, but does not fix these unambiguously. 
This ambiguity is resolved by the latter principles which
match the explicit singularity structure of the new antenna functions onto a specific antenna-subtraction scheme.

\subsection{Generic design principles}
\label{sec:generalprinciples}

The generic design principles outlined in Ref.~\cite{paper2} are sufficient to ensure that the antenna has the correct infrared limits.   Specifically, we impose the following requirements:
\begin{enumerate}[I.]
\item each antenna function has exactly two hard particles (``radiators'') which cannot become unresolved;
\item each antenna function captures all (multi-)soft limits of its unresolved particles;
\item where appropriate (multi-)collinear and mixed soft and collinear limits are decomposed over ``neighbouring'' antennae;
\item antenna functions do not contain any spurious (unphysical) limits;
\item antenna functions only contain singular factors corresponding to physical propagators; and
\item where appropriate, antenna functions obey physical symmetry relations (such as line reversal).
\end{enumerate}
As mentioned earlier, the original NNLO antenna functions derived in \cite{Gehrmann-DeRidder:2005svg,Gehrmann-DeRidder:2005alt,Gehrmann-DeRidder:2005btv} do not obey all of these requirements, as they typically violate (some of) these principles. This is particularly the case for quark-gluon or gluon-gluon antennae because the matrix elements they are derived from will inevitably have a divergent limit when one of the gluonic radiators becomes soft (thereby violating principle I). 

These principles will form the core of the algorithm for constructing  $X_{n+2}^\ell$ antennae with the desired infrared limits.

\subsection{Antenna-scheme-dependent design principles}
\label{sec:matching}

The generic principles are sufficient to produce compact analytic expressions that correctly capture the unresolved behaviour of $\ell$-loop matrix elements in the (multi-)soft and (multi-) collinear limits. Unlike the $\ell=0$ case, these unresolved limits have explicit singularities, and therefore the $X_{n+2}^\ell$ antennae constructed from them will also carry explicit $\e$-poles.  

However, it is straightforward to find terms that contain explicit singularities but which do not contribute in any of the unresolved regions.  Such terms, can be added to $X_{n+2}^\ell$ without violating any of the generic principles.  However, doing so will clearly change the explicit pole structure.  This means that the generic principles alone lead to an inherent ambiguity in defining the $X_{n+2}^\ell$ antennae.

If one wishes to design a full subtraction scheme, the real-virtual antenna must {\bf both} have the correct unresolved limits {\bf and} have an explicit pole structure of the correct form that cancels against other terms in the subtraction scheme.  Therefore, we need to resolve the ambiguity in the explicit $\e$ singularities by {\bf matching} onto a set of {\bf target} $\e$-pole structures that ensure that the subtraction terms in each multiplicity layer (a) correctly describe the unresolved limits of the matrix elements, and (b) precisely cancel the $\e$ singularities of the matrix elements.

To match onto a particular antenna-subtraction scheme, we therefore introduce one further principle: 
\begin{enumerate}
\item[VII.] where appropriate, combinations of terms that are not singular in the relevant unresolved regions can be added to match onto ``target poles'', $T(i_1^h,i_3,...,i_{n+2},i_2^h)$.
\end{enumerate}
To illustrate this principle for the case of the $X_3^1$, ``target poles'', $T(i^h,j,k^h)$, take the following schematic form within the NNLO antenna-subtraction scheme:
\begin{equation}
    T(i^h,j,k^h) = \frac{1}{\e^2} \left( \sum_{s} \left(\frac{s}{\mu^2}\right)^{-\e} \right) X_3^0(i^h,j,k^h),
\end{equation}
In order to match onto such ``target poles'', we are free to add certain combinations of terms. An example of such a combination of terms is, 
\begin{equation}
    \frac{1}{\e^2}\mu^{2\e}\left( s_{ik}^{-\e} + s_{ijk}^{-\e} - (s_{ij}+s_{ik})^{-\e} - (s_{ik}+s_{jk})^{-\e}\right) X_3^0 (i^h,j,k^h).
\end{equation}
which is not divergent in the soft $j$, collinear $ij$ and collinear $jk$ limits. 
It therefore does not affect the behaviour of the antenna function in those limits.
However, adding such a term clearly affects the explicit poles in the $X_3^1$ antenna as can be seen from the expansion in $\e$,
\begin{equation}
 \frac{1}{\e}\ln\left(1+\frac{s_{ij}s_{jk}}{s_{ijk}s_{ik}}\right) X_3^0 (i^h,j,k^h) +\order{\e^0}.
\end{equation}
This allows us to match the pole structure of the $X_3^1$ antenna to the other subtraction terms in a way that cancels the explicit poles at the RV level.

These seven principles are sufficient to devise an algorithm for constructing a general $X_{n+2}^\ell$ antenna function and here we will apply it to the construction of $X_3^1$ antenna functions with final-final kinematics. We will build the $X_3^1$ antenna functions from the infrared limits and match them to the NNLO antenna-subtraction scheme. The new $X_3^1$ antenna functions form the final ingredients for improved final-final antenna-subtraction at NNLO (along with the results of Ref.~\cite{paper2}). To test for the consistency of these ingredients, one has to integrate the real-virtual antenna over the antenna phase space, and combine all the various integrated implicit singularities to cancel the explicit singularities of the two-loop matrix elements. This is detailed in full in Section~\ref{sec:J22}.

\section{The algorithm}
\label{sec:algorithm}

In Ref.~\cite{paper2}, we proposed a general algorithm to build (multiple-)real radiation antenna functions at tree-level. In this paper, we extend this algorithm to the construction of $X_{n+2}^\ell$ antenna functions, where $\ell \neq 0$.

Unlike the algorithm for real-radiation antenna functions, the algorithm for $X_{n+2}^\ell$ antenna functions has two distinct stages:
\begin{description}
    \item[Stage 1] In this step we ensure that the antenna function has the correct infrared singular limits.
    This stage closely follows the algorithm for real-radiation antennae in Ref.~\cite{paper2}.
    We systematically start from the most singular limit, and build the list of target functions, $\{L_i\}$, from relevant (multi-)soft, (multi-)collinear, and soft-collinear limits. 
    
    As in Ref.~\cite{paper2}, we define a down- ($\PPdown_i$) and up-projector ($\PPup_i$) for each unresolved limit ($L_i$) to be included. A down-projector $\PPdown_i$ maps the invariants of the full phase space to the relevant subspace. 
    An associated up-projector $\PPup_i$ restores the full phase space by re-expressing all variables valid in the sub-space in terms of invariants valid in the full phase space. 
    It is to be emphasised that down-projectors $\PPdown_i$ and up-projectors $\PPup_i$ are typically not inverse to each other, as down-projectors destroy information about less-singular and finite pieces.

    The down-projectors are necessary to identify the overlapping region between the antenna function developed so far and the target function associated with the unresolved limit under consideration. Conversely, up-projectors express the argument in terms of antenna invariants. Furthermore, through careful selection of the up-projectors, the antenna function can be exclusively represented using invariants corresponding to physical propagators.

    The set of target functions provides a clear definition of the antenna function's behavior in all unresolved limits specific to the particular antenna being considered. In each unresolved limit, the antenna function must approach the corresponding target function to accurately capture the singular behavior exhibited by the squared matrix element. Additionally, the antenna function must remain finite in all limits not explicitly described by a target function. This crucial aspect guarantees the absence of spurious singularities (unlike antenna functions extracted directly from physical matrix elements).

    As explained in Ref.~\cite{paper2}, the algorithm, which ensures the above characteristics and meets the generic design principles, can be written as 
    \begin{equation}
    \label{eq:algorithm}
    \begin{split}
    X^\ell_{n+2;1} &= \PPup_1 L_1 \, , \\
    X^\ell_{n+2;2} &= X^\ell_{n+2;1} + \PPup_2 (L_2 - \PPdown_2 X^\ell_{n+2;1}) \, ,\\
    & \vdots \\
    X^\ell_{n+2;N} &= X^\ell_{n+2;N-1} + \PPup_N (L_N-\PPdown_N X^\ell_{n+2;N-1}) \, ,
    \end{split}
    \end{equation}
    where $X^\ell_{n+2;N}$ is the output of {\bf Stage 1}. 

    \item[Stage 2] The output of {\bf Stage 1} guarantees that $X^\ell_{n+2;N}$ has the chosen unresolved limits $\{L_i\}$.
    However, as discussed above this does not uniquely determine the mixed real-virtual antenna since one can construct a term (which we will denote by $\FinPoles$) which contains poles in $\e$ but does not contribute in any of the unresolved limits. One is therefore at liberty to define different antenna-subtraction schemes that differ by explicit $\e$-singular terms that do not affect the unresolved singular limits of the antenna.  
    
    We therefore add an antenna-scheme-dependent {\bf Stage 2} that ensures that the $X_{n+2}^\ell$ antenna has the correct explicit poles to {\bf match} onto the other types of subtraction terms in the desired antenna-subtraction scheme. We {\bf fix}  the scheme by 
    specifying that the explicit $\e$-poles, match certain defined ``target poles'', 
    \begin{equation}
        T = T(i_1^h,i_3,...,i_{n+2},i_2^h).
    \end{equation} 
    These target poles must be selected such that the constructed $X_{n+2}^\ell$ is more convenient for use in a wider $\text{N}^{n+\ell} \text{LO}$ subtraction scheme. Different schemes would entail different choices for $T$.  
   
    As in {\bf Stage 1}, we introduce certain projectors $\PPdown_T,\PPup_T$ (at the relevant perturbative order) to identify these additional $\e$-singular contributions which meet all the design principles. Schematically, we can write this final step of the algorithm as
    \begin{equation}
    \label{eq:Xfinal}
        X^\ell_{n+2} \equiv X^\ell_{n+2;N} + \PPup_T (T-\PPdown_T X^\ell_{n+2;N}).
    \end{equation}
    and we require
    \begin{equation}
        \PPdown_i \PPup_T (T-\PPdown_T X^\ell_{n+2;N}) \equiv 0 \qquad \forall \hspace{0.2cm} i=1,..,N.
    \end{equation}
    For later convenience we define the contribution from {\bf Stage 2} to be,
    \begin{equation}
    \FinPoles \equiv \PPup_T (T-\PPdown_T X^\ell_{n+2;N}).
    \end{equation}
    
\end{description}

Taking into account both {\bf Stage 1} and {\bf Stage 2}, the constructed mixed real-virtual antenna for a given set of infrared limits $\{L_i\}$ and matched to a scheme in which the required $\e$-poles are defined by $T$,  will satisfy
    \begin{align}
    \PPdown_i X^\ell_{n+2} &\equiv  L_i \qquad \forall  \hspace{0.2cm} i=1,..,N \, , \\
    \PPdown_T X^\ell_{n+2} &\equiv T.
    \end{align}
    
\section{Construction of real-virtual antenna functions}
\label{sec:X31construction}

The above design principles and algorithm have been set-out for the construction of a general $X^\ell_{n+2}$ antenna function. Now we specialise to the case of constructing real-virtual $X_3^1$ antenna functions. Together with the new $X_3^0$ and $\X$ of Ref.~\cite{paper2}, the $X_3^1$ functions complete the re-formulation of all antenna functions necessary for NNLO calculations, which now meet the design principles. 

We demonstrate the construction of real-virtual antenna functions $X_3^1(i^h_a,j_b,k^h_c)$, where the particle types are denoted by $a$, $b$, and $c$, which carry four-momenta $i$, $j$, and $k$, respectively. Particles $a$ and $c$ should be hard, and the antenna functions must have the correct limits when particle $b$ is unresolved. Frequently, we drop explicit reference to the particle labels in favour of a specific choice of $X$ according to Table~\ref{tab:X31}.

\begin{table}[t]
\centering
\begin{tabular}{ccc}
\underline{Quark-antiquark} & & \\
$qg\bar{q}$ & $X_3^1(i_q^h,j_g,k_{\bar{q}}^h)$ & $\Arv(i^h,j,k^h)$ \\
 & $\widetilde{X}_3^1(i_q^h,j_g,k_{\bar{q}}^h)$ & $\Atrv(i^h,j,k^h)$ \\
 & $\widehat{X}_3^1(i_q^h,j_g,k_{\bar{q}}^h)$ & $\Ahrv(i^h,j,k^h)$ \\
\underline{Quark-gluon} & &  \\
$qgg$ & $X_3^1(i_q^h,j_g,k_g^h)$ & $\Drv(i^h,j,k^h)$  \\
 & $\widetilde{X}_3^1(i_q^h,j_g,k_g^h)$ & $\Dtrv(i^h,j,k^h)$  \\
 & $\widehat{X}_3^1(i_q^h,j_g,k_g^h)$ & $\Dhrv(i^h,j,k^h)$  \\
$q\bar{Q}Q$ & $X_3^1(i_q^h,j_{\bar{Q}},k_Q^h)$  & $\Erv(i^h,j,k^h)$ \\ 
 & $\widetilde{X}_3^1(i_q^h,j_{\bar{Q}},k_Q^h)$  & $\Etrv(i^h,j,k^h)$ \\ 
  & $\widehat{X}_3^1(i_q^h,j_{\bar{Q}},k_Q^h)$  & $\Ehrv(i^h,j,k^h)$ \\ 
\underline{Gluon-gluon} & & \\
$ggg$ & $X_3^1(i_g^h,j_g,k_g^h)$ & $\Frv(i^h,j,k^h)$  \\
 & $\widehat{X}_3^1(i_g^h,j_g,k_g^h)$ & $\Fhrv(i^h,j,k^h)$  \\
$g\bar{Q}Q$ & $X_3^1(i_g^h,j_{\bar{Q}},k_Q^h)$  & $\Grv(i^h,j,k^h)$ \\ 
 & $\widetilde{X}_3^1(i_g^h,j_{\bar{Q}},k_Q^h)$  & $\Gtrv(i^h,j,k^h)$ \\ 
  & $\widehat{X}_3^1(i_g^h,j_{\bar{Q}},k_Q^h)$  & $\Ghrv(i^h,j,k^h)$ \\ 
\end{tabular}
\caption{Identification of $X_3^1$ antenna according to the particle type and colour-structures. These antennae only contain singular limits when particle $b$ (or equivalently momentum $j$) is unresolved, in addition to explicit $\e$ poles. Antennae are classified as quark-antiquark, quark-gluon and gluon-gluon according to the particle type of the parents (i.e. after the antenna mapping). }
\label{tab:X31}
\end{table}

For the specific case of $X_3^1(i^h_a,j_b,k^h_c)$ there are three such limits (meeting the generic design principles), corresponding to particle $b$ becoming soft, particles $a$ and $b$ becoming collinear, or particles $c$ and $b$ becoming collinear, so that the list of target functions is,
    \begin{equation}
    \begin{split}
        L_1(i^h,j,k^h) &= S_b^{(1)}(i^h,j,k^h) \, , \\
        L_2(i^h,j,k^h) &= P^{(1)}_{ab}(i^h,j) \, , \\
        L_3(i^h,j,k^h) &= P^{(1)}_{cb}(k^h,j) \, .
    \end{split}
    \label{eq:RVL}
    \end{equation}
The precise definitions of the one-loop soft factor $S_b^{(1)}$ and the one-loop splitting functions $P^{(1)}_{ab}$ are well known and we organise them in our notation in Section~\ref{sec:limits}.

In addition, in order to match onto a particular antenna-subtraction scheme, we require a target pole structure, $T \equiv T(i^h,j,k^h)$.  
We want to match the constructed $X_3^1$ to the full NNLO antenna-subtraction scheme, we therefore require the $\e$-poles to have a similar $\e$-pole structure to Ref.~\cite{Gehrmann-DeRidder:2005btv}. This means a collection of $\e$-poles multiplying $X_3^0$ antennae. 
By removing the contribution to the poles from the renormalisation term, we write the full set of target poles, $T(i^h,j,k^h)$, for the unrenormalised $X_3^1$ as, 
\begin{eqnarray}
\label{eq:targetA}
    T(i_q^h,j_g,k_{\bar{q}}^h)  &=&  -\frac{\Af}{\e^2} \left(S_{ij} + S_{jk} - S_{ijk}  \right) A_3^0 (i^h,j,k^h), \\ 
    \widetilde{T}(i_q^h,j_g,k_{\bar{q}}^h) &=&  
    -\frac{\Af}{\e^2} \left(S_{ijk} - S_{ik} \right) A_3^0 (i^h,j,k^h), \\ 
     \widehat{T}(i_q^h,j_g,k_{\bar{q}}^h) &=&  0, \\ 
     T(i_q^h,j_g,k_{g}^h) &=& -\frac{\Af}{\e^2} \left(S_{ij} + S_{[ik+jk]} + S_{jk} -2 S_{ijk} \right) D_3^{0} (i^h,j,k^h), \\ 
     \widetilde{T}(i_q^h,j_g,k_{g}^h) &=& -\frac{\Af}{\e^2} \left(S_{ik} - S_{[ik+jk]}  \right) D_3^{0} (i^h,j,k^h), \\ 
     \widehat{T}(i_q^h,j_g,k_{g}^h) &=&   0, \\ 
     T(i_q^h,j_{\bar{Q}},k_{Q}^h) &=&  -\Af\left[\frac{1}{\e^2} \left(S_{ij} + S_{ik} - 2 S_{ijk}   \right) - \frac{13}{6\e}S_{jk} \right] E_3^0 (i^h,j,k^h), \\ 
     \widetilde{T}(i_q^h,j_{\bar{Q}},k_{Q}^h) &=&  - \Af \left( \frac{1}{\e^2} + \frac{3}{2 \e} \right) S_{jk} E_3^0 (i^h,j,k^h), \\ 
     \widehat{T}(i_q^h,j_{\bar{Q}},k_{Q}^h) &=& - \Af \frac{2}{3 \e} S_{jk} E_3^0 (i^h,j,k^h), \\ 
     T(i_g^h,j_g,k_{g}^h) &=&  -\frac{\Af}{\e^2} \left(S_{ij} + S_{ik} + S_{jk} -2 S_{ijk} \right) F_3^0 (i^h,j,k^h), \\ 
     \widehat{T}(i_g^h,j_g,k_{g}^h) &=&  0 , \\ 
     T(i_g^h,j_{\bar{Q}},k_{Q}^h) &=&  -\Af\left[\frac{1}{\e^2} \left(S_{ij} + S_{ik} - 2 S_{ijk}   \right) - \frac{13}{6\e}S_{jk} \right] G_3^0 (i^h,j,k^h), \\ 
     \widetilde{T}(i_g^h,j_{\bar{Q}},k_{Q}^h) &=&  - \Af \left( \frac{1}{\e^2} + \frac{3}{2 \e} \right) S_{jk} G_3^0 (i^h,j,k^h), \\
\label{eq:targetG}
\widehat{T}(i_g^h,j_{\bar{Q}},k_{Q}^h) &=& - \Af \frac{2}{3 \e} S_{jk} G_3^0 (i^h,j,k^h).
\end{eqnarray}
Here $\Af$ is an overall factor defined as
\begin{equation}
\label{eq:Afdef}
    \Af = e^{\e \gamma} \frac{\Gamma^2(1-\e) \Gamma(1+\e)}{\Gamma(1-2\e)} \text{Re} (-1)^{-\e} .
\end{equation}
This factor ensures that the $X_3^1$ antennae derived here have the same overall normalisation as those in Ref.~\cite{Gehrmann-DeRidder:2005btv}. We have also introduced the convenient notation to separate the loop-type structures from the unresolved-type structures: \begin{equation}
\label{eq:Sdefs}
    S_{ij} = \left(\frac{s_{ij}}{\mu^2}\right)^{-\e},\qquad
    S_{ijk} = \left(\frac{s_{ijk}}{\mu^2}\right)^{-\e}, \qquad 
    S_{[ik+jk]} = \left(\frac{s_{ik}+s_{jk}}{\mu^2}\right)^{-\e}.
\end{equation}

The $X_3^0 (i^h,j,k^h)$ antennae appearing in Eqs.~\eqref{eq:targetA}--\eqref{eq:targetG} are those constructed in Ref.~\cite{paper2}. Therefore, there are some differences compared to the pole structure 
in Ref.~\cite{Gehrmann-DeRidder:2005btv} precisely due to differences between the definitions of the $X_3^0 (i^h,j,k^h)$ of Ref.~\cite{paper2} and the $X_3^{0,\text{OLD}} (i^h,j,k^h)$ of Ref.~\cite{Gehrmann-DeRidder:2005btv} (see Ref.~\cite{paper2} for a full discussion). 
This is particularly the case for the $D_3^0$ and $F_3^0$ antennae where the unresolved singularities of $F_3^{0,\text{OLD}}$ are assigned to three
$F_3^{0}$ antenna, while the unresolved singularities present in 
$D_3^{0,\text{OLD}}$ are found by combining two $D_3^{0}$ antenna.   

Finally, we note that the $\e$-singularities of 
$D_3^{1,\text{OLD}}$ of Ref.~\cite{Gehrmann-DeRidder:2005btv} are split between $\Drv$ and a new type of antenna, $\Dtrv$,
\begin{equation}
   \PPdown_T D_3^{1,\text{OLD}} \sim T(i_q^h,j_g,k_{g}^h) +\widetilde{T}(i_q^h,j_g,k_{g}^h) + ( j \leftrightarrow k) ,
\end{equation}
where $\sim$ reflects the fact that the LHS multiplies $D_3^{0,\text{OLD}}$ while the RHS multiplies $D_3^{0}$.

\subsection{Template antennae}
For convenience, we define a general unrenormalised real-virtual antenna function in terms of the contributions produced by the various steps of the algorithm.  At leading-colour, we have,
\begin{eqnarray}
\label{eq:X31def}
X_3^1(i^h,j,k^h) &=& \Ssoft^{(1)}(i^h,j,k^h) + \Scol^{(1)}(i^h,j;k^h) + \Scol^{(1)}(k^h,j;i^h) \\ \nonumber
&&+  \FinPoles(i^h,j,k^h) \, ,
\end{eqnarray}
while the corresponding sub-leading-colour expression is,
\begin{eqnarray}
\label{eq:X31tdef}
\widetilde{X}_3^1(i^h,j,k^h) &=& \Scolt^{(1)}(i^h,j;k^h) + \Scolt^{(1)}(k^h,j;i^h) 
\\ \nonumber
&&+  \FinPolest(i^h,j,k^h) \, ,
\end{eqnarray}
and the quark-loop contribution is,
\begin{eqnarray}
\label{eq:X31hdef}
\widehat{X}_3^1(i^h,j,k^h) &=& \Scolh^{(1)}(i^h,j;k^h) + \Scolh^{(1)}(k^h,j;i^h)\\ \nonumber
&&+  \FinPolesh(i^h,j,k^h) \, ,
\end{eqnarray}
since the one-loop soft factor is only non-zero at leading-colour.
The meanings of the individual terms in Eqs.~\eqref{eq:X31def}--\eqref{eq:X31hdef} will be made clear in the following subsections, however, we note that in each equation, the first line is produced by {\bf Stage 1} of the algorithm, and the second line is added in {\bf Stage 2}.  Therefore, we expect that 
\begin{align}
    \PSdown_j \FinPoles(i^h,j,k^h) &= 0 \, , \\
    \PCdown_{ij} \FinPoles(i^h,j,k^h) &= 0 \, , \\
    \PCdown_{jk} \FinPoles(i^h,j,k^h) &= 0 \, ,   
\end{align}
and similarly for $\FinPolest(i^h,j,k^h)$ and $\FinPolesh(i^h,j,k^h)$.

\subsection{Stage 1}

All $X_3^1 (i^h,j,k^h)$ antenna functions are defined over the full three-particle phase space, whereas each unresolved limit lives on a restricted part of phase space: the $j$ soft limit, the $ij$ collinear limit, and the $jk$ collinear limit. 

We define the soft down-projector by its action on integer powers of invariants as
\begin{equation}
    \PSdown_j: \begin{cases} 
    s_{ij} \mapsto \lambda s_{ij}, 
    s_{jk} \mapsto \lambda s_{jk}, \\
    s_{ijk} \mapsto s_{ik},
    \end{cases}
\end{equation}
and keep only the terms proportional to $\lambda^{-2}$. 

For the corresponding up-projector $\PSup_j$ we choose a trivial mapping which leaves all variables unchanged. The collinear down-projector acts on integer powers of invariants and is defined in analogy to Eq.~(4.16) of Ref.~\cite{paper2},
\begin{equation}
    \PCdown_{ij}: \begin{cases}
    s_{ij} \mapsto \lambda s_{ij}, \\
    s_{ik} \mapsto \omxj (s_{ik}+s_{jk}),
    s_{jk} \mapsto \xj (s_{ik}+s_{jk}), 
    s_{ijk} \mapsto s_{ik} + s_{jk},
    \end{cases}
\end{equation}
but keeps only terms of order $\lambda^{-1}$.  The corresponding up-projector is the same as in Eq.~(4.17) of \cite{paper2},
\begin{equation}
    \PCup_{ij}: \begin{cases}
    \xj \mapsto s_{jk}/s_{ijk},
    \omxj \mapsto s_{ik}/s_{ijk}, \\
    s_{ik}+s_{jk} \mapsto s_{ijk}.
    \end{cases} \, 
\end{equation}
This up-projector ensures the presence of $s_{ijk}$ denominators, which are present in matrix elements corresponding to physical propagators and means that the same integration tools for one-loop matrix elements can be used in the integration of the constructed $X_3^1$ over its Lorentz-invariant antenna phase space.

The subtracted single-unresolved one-loop factors are built from unrenormalised colour-ordered limits and are given by
\begin{align}
    \Ssoft^{(1)}(i^h,j,k^h) &= \PSup_j S_b^{(1)}(i^h,j,k^h) \, ,\\
    \Scol^{(1)}(i^h,j;k^h) &= \PCup_{ij}\left(P_{ab}^{(1)}(i^h,j) - \PCdown_{ij}\Ssoft^{(1)}(i^h,j,k^h) \right) \, , \\
    \Scol^{(1)}(k^h,j;i^h) &= \PCup_{kj}\left(P_{cb}^{(1)}(k^h,j) - \PCdown_{kj}\left(\Ssoft^{(1)}(i^h,j,k^h) + \Scol^{(1)}(i^h,j;k^h) \right) \right) \, , \nonumber \\
    &\equiv \PCup_{kj}\left(P_{cb}^{(1)}(k^h,j) - \PCdown_{kj}\Ssoft^{(1)}(i^h,j,k^h)  \right) \, ,
\end{align}
and analogously for the sub-leading colour and quark-flavour contributions. The subscripts $a,b,c$ represent the particle types which carry momenta $i,j,k$ respectively. The unrenormalised one-loop single-unresolved limits are listed in full in Section~\ref{sec:limits}. We have used the feature that the only overlap between the $ij$- and $jk$-collinear limits occurs in $\Ssoft^{(1)}$ so that
\begin{equation}
    \PCdown_{kj}\Scol^{(1)}(i^h,j;k^h) = 0,
\end{equation}
which was also observed in Ref.~\cite{paper2}. 

At this point, we have iteratively constructed the quantity 
\begin{equation}
\label{eq:X33}
    X_{3;3}^1(i^h,j,k^h) = 
    \Ssoft^{(1)}(i^h,j,k^h)+
    \Scol^{(1)}(i^h,j;k^h)+
    \Scol^{(1)}(k^h,j;i^h),
\end{equation}
such that,
\begin{align}
    \PSdown_j X_{3;3}^1(i^h,j,k^h) &= S_b^{(1)}(i^h,j,k^h) \, , \\
    \PCdown_{ij} X_{3;3}^1(i^h,j,k^h) &= P_{ab}^{(1)}(i^h,j) \, , \\
    \PCdown_{jk} X_{3;3}^1(i^h,j,k^h) &= P_{cb}^{(1)}(k^h,j) \, ,   
\end{align}
which carries all of the desired unresolved limits.

\subsection{Stage 2}
\label{subsec:stage2}

We now turn to the construction of $\FinPoles(i^h,j,k^h)$, which
does not contribute to any unresolved limit but does contain explicit $\e$ poles. In the language of section~\ref{sec:algorithm}, then schematically
\begin{equation}
\FinPoles(i^h,j,k^h) 
\equiv \PPup_T \left(T(i^h,j,k^h)-\PPdown_T X^1_{3;3(i^h,j,k^h)}\right).
\end{equation}

We observe that each of the target pole structures in Eqs.~\eqref{eq:targetA}--\eqref{eq:targetG} is of the form,
\begin{equation}
    \Poles \times \XLO (i^h,j,k^h),
\end{equation}
where $\Poles$ is combination of $\e$-poles and factors like $s_{ij}^{-\e}$. 
We therefore choose to achieve {\bf Stage 2} through two iterative steps (rather than one), adding a projector for each step:
\begin{description}
\item[Step 1] We introduce a projector $\PXup$ (and the trivial projector $\PXdown$) to ensure that the $\e$-poles are proportional to $\XLO(i^h,j,k^h)$; and
\item[Step 2] We introduce projectors $\Peup$ and $\Pedown$ to adjust the pole structure multiplying $\XLO$ to match $T(i^h,j,k^h)$.
\end{description}

\subsection*{Step 1}
We define the projector $\PXup$ such that,
\begin{equation}
    \PXup: \begin{cases}
    P_{ab}^{(0)}(i^h,j) \mapsto X_3^0 (i^h,j,k^h),\\
    P_{cb}^{(0)}(k^h,j) \mapsto X_3^0 (i^h,j,k^h).
    \end{cases} 
\end{equation}
The inverse projector $\PXdown$ is simply unity.  

We define $\FinPoles_X(i^h,j,k^h)$ to be the contribution arising from the action of $\PXup$ such that,
\begin{align}
\label{eq:finpolesX}
    \FinPoles_X(i^h,j,k^h) &= (\PXup - 1) ( \Ssoft^{(1)}(i^h,j,k^h) + \Scol^{(1)}(i^h,j;k^h) + \Scol^{(1)}(k^h,j;i^h) ), \nonumber \\
    &= (\PXup - 1) X_{3;3}^1(i^h,j,k^h).
\end{align}
$\Ssoft^{(1)}(i^h,j,k^h)$ does not contain splitting functions, so the action of $\PXup$ on $\Ssoft$ is trivial,
\begin{equation}
   \PXup \Ssoft^{(1)}(i^h,j,k^h) \equiv \Ssoft^{(1)}(i^h,j,k^h), 
\end{equation}
which guarantees,
\begin{align}
\label{eq:noIRfinpolesX1}
    \PSdown_j \FinPoles_X(i^h,j,k^h) &= 0 \, .  
\end{align}
Furthermore, because of the structure of the one-loop splitting functions, we also have
\begin{align}
\label{eq:noIRfinpolesX2}
    \PCdown_{ij} \FinPoles_X(i^h,j,k^h) &= 0 \, , \nonumber \\
    \PCdown_{jk} \FinPoles_X(i^h,j,k^h) &= 0 \, .  
\end{align}
We will explain more clearly how this is achieved in the specific example of $A_3^1$ in Section~\ref{sec:A31construct}.
We also note that following the iterative structure of Eq.~\eqref{eq:algorithm}, we define the fourth step of the algorithm to be
\begin{align}
\label{eq:X34}
    X_{3;4}^1(i^h,j,k^h) &= X_{3;3}^1(i^h,j,k^h) + (\PXup-1) X_{3;3}^1(i^h,j,k^h), \\
              &\equiv \PXup X_{3;3}^1(i^h,j,k^h),
\end{align}
where $X_{3;3}^1(i^h,j,k^h)$ is given in Eq.~\eqref{eq:X33}.

\subsection*{Step 2}

The operators $\Pedown$ and $\Peup$ are defined as follows.

$\Pedown$ is defined by Laurent expanding the argument in $\e$ and discarding terms of $\order{\e^0}$ and higher. 
    
$\Peup$ is defined by extending the argument to an all-orders  expression in $\e$, which agrees with the argument up to $\order{\e^0}$. This is not a unique action. For the case of the $X_3^1$ we choose, where possible, for $\Peup$ to result in linear combinations of $\{ s_{ik}^{-\e}, s_{ijk}^{-\e}, (s_{ik} + s_{jk})^{-\e} \}$ (which are simply-integrable objects) multiplied by simple $\e$-poles and $X_3^0$.
Only two structures appear in the construction of the $X_3^1$:
\begin{equation}
   \Peup \frac{1}{\e} \ln \left(1 + \frac{s_{ij} s_{jk} }{s_{ik} s_{ijk}} \right)  = \frac{1}{\e^2} \Lambda_1(i^h,j,k^h) , 
\end{equation}
\begin{equation}
   \Peup \frac{2}{\e} \ln \left( 1- \frac{s_{jk} }{s_{ijk}} \right)  = \frac{2}{\e^2} \Lambda_2(i^h,j,k^h) , 
\end{equation}
where 
\begin{eqnarray}
    \Lambda_1(i^h,j,k^h) &=&  S_{ik} + S_{ijk} - S_{[ik+jk]} - S_{[ik+ij]} , \\
    \Lambda_2(i^h,j,k^h) &=& S_{ijk} - S_{[ik+ij]} ,
\end{eqnarray}
with $S_{ik}$ etc defined as in Eq.~\eqref{eq:Sdefs}.

We define $\FinPoles_{\e}(i^h,j,k^h)$ to be the contribution arising from the action of $\Peup$ and $\Pedown$ such that,
\begin{eqnarray}
\label{eq:finpolese}
    \FinPoles_{\e}(i^h,j,k^h) &=& \Peup \bigg( \Pedown T(i^h,j,k^h) - \Pedown \bigg[ \Ssoft^{(1)}(i^h,j,k^h) \\ \nonumber
    && + \Scol^{(1)}(i^h,j;k^h)  +  \Scol^{(1)}(k^h,j;i^h) +  \FinPoles_X (i^h,j,k^h) \bigg] \bigg).
\end{eqnarray}
These contributions typically contain a factor which suppresses all the unresolved limits in the $X_3^0$ to which it multiplies. $\Lambda_1$ suppresses any contributions to the soft-$j$ limit or the collinear-$ij$ or collinear-$jk$ limits. $\Lambda_2$ suppresses contributions to the soft-$j$ or collinear-$jk$ limits so that
\begin{align}
\label{eq:noIRfinpolese}
    \PSdown_j \FinPoles_{\e}(i^h,j,k^h) &= 0 \, ,\nonumber \\
    \PCdown_{ij} \FinPoles_{\e}(i^h,j,k^h) &= 0 \, , \nonumber\\
    \PCdown_{jk} \FinPoles_{\e}(i^h,j,k^h) &= 0 \, .  
\end{align}
We will explain more clearly how this works in detail in the specific example of $A_3^1$ in section~\ref{sec:A31construct}.

In the iterative language of Eq.~\eqref{eq:algorithm} the fifth and final step of the algorithm is thus
\begin{align}
    X_{3;5}^1(i^h,j,k^h) &=   X_{3;4}^1(i^h,j,k^h) +  \Peup \left(\Pedown T(i^h,j,k^h) - \Pedown X_{3;4}^1(i^h,j,k^h) \right),
\end{align}
where we define the complete constructed antenna function,
\begin{equation}
     X_3^1(i^h,j,k^h) \equiv X_{3;5}^1(i^h,j,k^h) .
\end{equation}

It is convenient to combine the contributions from Eqs.~\eqref{eq:finpolesX} and \eqref{eq:finpolese}, to obtain a single contribution (as in the antenna templates of Eqs.~\eqref{eq:X31def}, \eqref{eq:X31tdef} and \eqref{eq:X31hdef}), and we define
\begin{equation}
    \FinPoles(i^h,j,k^h) \equiv \FinPoles_X(i^h,j,k^h) + \FinPoles_{\e}(i^h,j,k^h).
\end{equation}
It is to be emphasised again that $\FinPoles(i^h,j,k^h)$ does not contribute in any unresolved limits, but does carry explicit poles in $\e$.
Indeed, using Eqs.~\eqref{eq:noIRfinpolesX1}, ~\eqref{eq:noIRfinpolesX2}  and \eqref{eq:noIRfinpolese}, it is straightforward to see that
\begin{align}
\label{eq:noIRfinpoles}
    \PSdown_j \FinPoles(i^h,j,k^h) &= 0 \, ,\nonumber \\
    \PCdown_{ij} \FinPoles(i^h,j,k^h) &= 0 \, , \nonumber\\
    \PCdown_{jk} \FinPoles(i^h,j,k^h) &= 0 \, .  
\end{align}

Finally, the algorithm of this paper ensures that
\begin{align}
    \PSdown_j X_3^1(i^h,j,k^h) &= S_b^{(1)}(i^h,j,k^h) \, , \\
    \PCdown_{ij} X_3^1(i^h,j,k^h) &= P_{ab}^{(1)}(i^h,j) \, , \\
    \PCdown_{jk} X_3^1(i^h,j,k^h) &= P_{cb}^{(1)}(k^h,j) \, ,   
\end{align}
and 
\begin{align}
    \Pedown X_3^1(i^h,j,k^h) &= \Pedown T(i^h,j,k^h)\, .
\end{align}

\subsection{Renormalisation}

As a final step, we renormalise the antennae at scale $\mu$, 
\begin{eqnarray}
X_3^1(i^h,j,k^h) &\mapsto & X_3^1(i^h,j,k^h) -  \frac{ \bz}{\e} X_3^0(i^h,j,k^h) \, , \\
\label{eq:X31defR}
\widetilde{X}_3^1(i^h,j,k^h) &\mapsto& \widetilde{X}_3^1(i^h,j,k^h) \, , \\
\label{eq:X31tdefR}
\widehat{X}_3^1(i^h,j,k^h) &\mapsto& \widehat{X}_3^1(i^h,j,k^h)-  \frac{ \bzf}{\e}  X_3^0(i^h,j,k^h) \, .
\label{eq:X31hdefR}
\end{eqnarray}
We use the colour decomposition of $\Bz$,
\begin{equation}
    \Bz = N_c \bz + N_F \bzf ,
\end{equation}
where $\bz = 11/6$ and $\bzf=-1/3$.

\subsection{$\Arv$ construction in full detail}
\label{sec:A31construct}

To make the construction explicit, we work through the construction of $A_3^1$ as an example before describing the full set of improved real-virtual antenna functions in Section~\ref{sec:X31}. 
$\Arv (i_q^h,j_g,k_{\bar{q}}^h)$ is the leading-colour antenna function with quark and antiquark hard radiators, which encapsulates the one-loop limits when the gluon becomes unresolved. The relevant unresolved limits are
\begin{equation}
    \Sgone(i^h,j,k^h), \qquad\Pqgone(i_q^h,j_g), \qquad\Pqgone(k_{\bar{q}},j_g) ,
\end{equation}
which are given in Section~\ref{sec:limits}.  
Additionally, we choose a target for the $\e$-poles before renormalisation, which is consistent with the above limits but matches the $\e$-pole structures appearing in the antenna-subtraction scheme. The target pole structure for $\Arv$ is given by
\begin{equation}
\label{eq:Atarget}
T(i_q^h,j_g,k_{\bar{q}}^h)  =  \frac{\Af}{\e^2} \left(S_{ij} + S_{jk} - S_{ijk}  \right) A_3^0 (i^h,j,k^h).
\end{equation}
We want to match the constructed $X_3^1$ to the full NNLO antenna-subtraction-scheme, we therefore require the $\e$-poles to have a similar $\e$-pole structure to Ref.~\cite{Gehrmann-DeRidder:2005btv}.

We choose to simplify our notation by introducing the following structure 
\begin{eqnarray}
\label{eq:Ghyperdef}
G(w,\e) &=&  {}_2F_1\left(1,\e,1+\e,-w\right) -1  , \nonumber \\
&=& - \sum_{n=1}^{\infty} (-\e)^n \Li_{n}\left(-w\right), \nonumber \\
&\equiv& (1+w)^{-\e} {}_2F_1\left(\e,\e,1+\e,\frac{w}{1+w}\right) -1  ,
\end{eqnarray}
where $w = s_{jk}/s_{ik}$. 
Note that in the $w \to 0$ limit, $G(w,\e)$ vanishes.  

Before renormalisation, $\Arv$ is built iteratively in pieces in the following order:
\begin{eqnarray}
\Arv(i_q^h,j_g,k_{\bar{q}}^h) &=& \Ssoft^{(1)}(i_q^h,j_g,k_{\bar{q}}^h) + \Scol^{(1)}(i_q^h,j_g;k_{\bar{q}}^h) + \Scol^{(1)}(k_{\bar{q}}^h,j_g;i_q^h) \\ \nonumber
&&+  \FinPoles(i_q^h,j_g,k_{\bar{q}}^h) \, ,
\label{eq:A31def}
\end{eqnarray}
with the $\FinPoles$ contribution constructed in two steps as in Section~\ref{subsec:stage2} according to Eqs.~\eqref{eq:finpolesX} and \eqref{eq:finpolese},
\begin{equation}
\FinPoles(i_q^h,j_g,k_{\bar{q}}^h) 
= \FinPoles_X(i_q^h,j_g,k_{\bar{q}}^h) 
+  \FinPoles_{\e}(i_q^h,j_g,k_{\bar{q}}^h). 
\end{equation}

The first contribution is simply the one-loop soft factor,
\begin{equation}
\label{eq:SsoftA31}
    \Ssoft^{(1)}(i_q^h,j_g,k_{\bar{q}}^h) = \PSup_j \Sgone(i^h,j,k^h)  = - \Af\frac{\Gamma(1-\e)\Gamma(1+\e)}{\e^2} \frac{S_{ij}S_{jk}}{S_{ik}}  \Sgzero (i^h,j,k^h), 
\end{equation}
where $\Sgzero$ is the tree-level eikonal factor given in Appendix~\ref{app:treelimits}. 

The second piece is given by the overlap of the one-loop splittting function $\Pqgone(i^h,j)$ and the one-loop soft factor $\Ssoft^{(1)}$ in the $ij$ collinear limit, projected-up into the full phase-space: 
\begin{eqnarray}
\lefteqn{     \Scol^{(1)}(i_q^h,j_g;k_{\bar{q}}^h) } \nonumber \\
&=& \PCup_{ij}\left(\Pqgone(i^h,j) - \PCdown_{ij}\Ssoft^{(1)}(i_q^h,j_g,k_{\bar{q}}^h) \right), \nonumber \\ 
     &=&  \frac{\Af}{\e^2} \bigg[ -\Gamma(1-\e)\Gamma(1+\e) \frac{S_{ij}S_{jk}}{S_{ik}} \frac{\ome s_{jk}}{s_{ij}s_{ijk}} + S_{ij} G\left(\frac{s_{jk}}{s_{ik}},\e \right) \Pqgzero(i_q^h,j_g;k_{\bar{q}}^h)  \bigg]  \nonumber \\ 
    \label{eq:ScolA31a}
     &&+ S_{ij} \Af \frac{(s_{ijk}-\e s_{jk} )}{s_{ij}s_{ijk}} \frac{1}{2 (1-2\e)} . 
\end{eqnarray}
Here we use the short-hand notation 
\begin{equation}
  P_{ab}^{(n)} (i^h,j;k) = \PCup_{ij} P_{ab}^{(n)} (i^h,j), 
\end{equation}
to indicate an $n$-loop splitting function up-projected into the full phase space of the antenna and the tree-level splitting functions, $P_{ab}^{(0)}$ are given in Appendix~\ref{app:treelimits}.

The third contribution is given by
\begin{eqnarray}
\lefteqn{\Scol^{(1)}(k_{\bar{q}}^h,j_g;i_q^h)} \nonumber \\
&=& \PCup_{kj}\left(\Pqgone(k_q^h,j_g) - \PCdown_{kj}\left(\Ssoft^{(1)}(i_q^h,j_g,k_{\bar{q}}^h) + \Scol^{(1)}(i_q^h,j_g;k_{\bar{q}}^h) \right) \right), \nonumber \\
&=& \PCup_{kj}\left(\Pqgone(k_q^h,j_g) - \PCdown_{kj}\Ssoft^{(1)}(i_q^h,j_g,k_{\bar{q}}^h)  \right) , \nonumber \\ 
    &=& \frac{\Af}{\e^2} \bigg[ -\Gamma(1-\e)\Gamma(1+\e) \frac{S_{ij}S_{jk}}{S_{ik}} \frac{\ome s_{ij}}{s_{jk}s_{ijk}} + S_{jk} G\left(\frac{s_{ij}}{s_{ik}},\e \right) \Pqgzero(k_{\bar{q}}^h,j_g;i_{q}^h)  \bigg] \nonumber \\ 
    \label{eq:ScolA31b}
     &&+  S_{jk} \Af \frac{(s_{ijk}-\e s_{ij} )}{s_{jk}s_{ijk}}\frac{1}{2 (1-2\e)}  . 
\end{eqnarray}
Recalling from Ref.~\cite{paper2} that 
\begin{eqnarray}
    A_3^0 (i_q^h,j_g,k_{\bar{q}}^h) &\equiv& \Sgzero (i^h,j,k^h) + \frac{\ome s_{jk}}{s_{ij}s_{ijk}} + \frac{\ome s_{ij}}{s_{jk}s_{ijk}}, \\ \nonumber
    &\equiv& \Pqgzero(i_q^h,j_g;k_{\bar{q}}^h) + \frac{\ome s_{ij}}{s_{jk}s_{ijk}} ,\\ \nonumber
    &\equiv& \Pqgzero(k_{\bar{q}}^h,j_g;i_{q}^h) + \frac{\ome s_{jk}}{s_{ij}s_{ijk}} ,
\end{eqnarray}
it is straightforward to see that the terms proportional to $S_{ij}S_{jk}/S_{ik}$ in Eqs.~\eqref{eq:SsoftA31},~\eqref{eq:ScolA31a} and \eqref{eq:ScolA31b}, combine to give a term which factorises onto $A_3^0(i_q^h,j_g,k_{\bar{q}}^h)$ such that 
\begin{eqnarray}
\label{eqn:runningA31}
\lefteqn{   \Ssoft^{(1)}(i_q^h,j_g,k_{\bar{q}}^h) + \Scol^{(1)}(i_q^h,j_g;k_{\bar{q}}^h) + \Scol^{(1)}(k_{\bar{q}}^h,j_g;i_q^h) } \nonumber \\ 
   &=&+ \frac{\Af}{\e^2}\bigg[ -\Gamma(1-\e)\Gamma(1+\e) \frac{S_{ij}S_{jk}}{S_{ik}} A_3^0(i^h,j,k^h) \nonumber \\ 
   &&+S_{ij} G\left(\frac{s_{jk}}{s_{ik}},\e \right) \Pqgzero(i_q^h,j_g;k_{\bar{q}}^h) + S_{jk}G\left(\frac{s_{ij}}{s_{ik}},\e \right) \Pqgzero(k_{\bar{q}}^h,j_g;i_{q}^h) \bigg] \nonumber \\ 
    &&  + S_{ij} \Af \frac{(s_{ijk}-\e s_{jk} )}{s_{ij}s_{ijk}} \frac{1}{2 (1-2\e)} 
    +  S_{jk} \Af \frac{(s_{ijk}-\e s_{ij} )}{s_{jk}s_{ijk}}\frac{1}{2 (1-2\e)}  .
\end{eqnarray}
This combination completes Stage 1 of the algorithm and is to some extent a complete construction of $\Arv$, in the sense that it encapsulates the fundamental one-loop unresolved limits we require of it.

{\bf Stage 2} of the algorithm preserves the unresolved limits but includes explicit poles that do not contribute in these limits.  The next piece $\FinPoles_X$ ensures that the explicit pole structure of the $\Arv$ factors onto $A_3^0$:
\begin{align}
    \FinPoles_X(i_q^h,j_g,k_{\bar{q}}^h) &= (\PXup - 1) \left( \Ssoft^{(1)}(i_q^h,j_g,k_{\bar{q}}^h) + \Scol^{(1)}(i_q^h,j_g;k_{\bar{q}}^h) + \Scol^{(1)}(k_{\bar{q}}^h,j_g;i_q^h) \right), \nonumber \\ 
    &= \frac{\Af}{\e^2} \bigg[\hspace{0.3cm} S_{ij} G\left(\frac{s_{jk}}{s_{ik}},\e \right) \left(A_3^0 (i_q^h,j_g,k_{\bar{q}}^h) - \Pqgzero(i_q^h,j_g;k_{\bar{q}}^h) \right) \nonumber \\ 
    &\hspace{1cm} + S_{jk}G\left(\frac{s_{ij}}{s_{ik}},\e \right) \left( A_3^0 (i_q^h,j_g,k_{\bar{q}}^h) - \Pqgzero(k_{\bar{q}}^h,j_g;i_{q}^h) \right) \bigg], \nonumber \\ 
    &= \frac{\Af}{\e^2} \bigg[ S_{ij} G\left(\frac{s_{jk}}{s_{ik}},\e \right) \frac{\ome s_{ij}}{s_{jk}s_{ijk}} + S_{jk}G\left(\frac{s_{ij}}{s_{ik}},\e \right) \frac{\ome s_{jk}}{s_{ij}s_{ijk}} \bigg].
    \label{eq:FinPolesA31A}
\end{align}
This term vanishes in the unresolved region for the following reason.  
The first term in the final line appears to have a singularity in the $jk$ collinear limit due to the $1/s_{jk}$ factor.  However, in this limit the hypergeometric function $G(s_{jk}/s_{ijk},\e)$ approaches zero  and this behaviour therefore suppresses the singularity due to the $1/s_{jk}$ factor.  A similar argument holds for the second term. As such, neither term in Eq.~\eqref{eq:FinPolesA31A} contributes to any unresolved limit, although they evidently do contribute explicit $\e$ poles. 
In summary,
\begin{align}
   & \PSdown_j \left( \frac{1}{s_{ij}s_{jk}} \times G\left(\frac{s_{jk}}{s_{ik}},\e \right) 
    \right) 
     \to  0 \, ,\nonumber \\
   & \PSdown_j  \left( \frac{1}{s_{ij}s_{jk}} \times G\left(\frac{s_{ij}}{s_{ik}},\e \right)\right)  \to  0 \, ,\nonumber \\
   & \PCdown_{ij}  \left( \frac{1}{s_{ij}} \times G\left(\frac{s_{ij}}{s_{ik}},\e \right) \right) \to  0 \, ,\nonumber \\
   & \PCdown_{jk} \left( \frac{1}{s_{jk}} \times  G\left(\frac{s_{jk}}{s_{ik}},\e \right) \right) \to  0 \, .\nonumber 
\end{align}

The running total for $\Arv$ is given by 
\begin{eqnarray}
\lefteqn{   \Ssoft^{(1)}(i_q^h,j_g,k_{\bar{q}}^h) + \Scol^{(1)}(i_q^h,j_g;k_{\bar{q}}^h) + \Scol^{(1)}(k_{\bar{q}}^h,j_g;i_q^h) + \FinPoles_X(i_q^h,j_g,k_{\bar{q}}^h) } \nonumber \\ 
   &=&+ \frac{\Af}{\e^2}\bigg[ -\Gamma(1-\e)\Gamma(1+\e) \frac{S_{ij}S_{jk}}{S_{ik}}   \\
   &&+S_{ij} G\left(\frac{s_{jk}}{s_{ik}},\e \right) + S_{jk}G\left(\frac{s_{ij}}{s_{ik}},\e \right) \bigg] A_3^0 (i_q^h,j_g,k_{\bar{q}}^h) , \nonumber \\ 
    &&  + S_{ij} \Af \frac{(s_{ijk}-\e s_{jk} )}{s_{ij}s_{ijk}} \frac{1}{2 (1-2\e)} 
    +  S_{jk} \Af \frac{(s_{ijk}-\e s_{ij} )}{s_{jk}s_{ijk}}\frac{1}{2 (1-2\e)}  .\nonumber
\end{eqnarray}
Effectively, the tree-level splitting functions in Eq.~\eqref{eqn:runningA31} have been promoted to full $A_3^0$ antenna functions.

The next contribution, $\FinPoles_{\e}$, is also part of antenna-scheme matching, for which we have the target pole structure proportional to $A_3^0$, given in Eq.~\eqref{eq:Atarget}. The resulting expression is given by
\begin{eqnarray}
    \FinPoles_{\e}(i_q^h,j_g,k_{\bar{q}}^h) &=& \Peup \bigg( \Pedown T(i_q^h,j_g,k_{\bar{q}}^h) - \Pedown \bigg[ \Ssoft^{(1)}(i_q^h,j_g,k_{\bar{q}}^h) \\ \nonumber
    && + \Scol^{(1)}(i_q^h,j_g;k_{\bar{q}}^h) + \Scol^{(1)}(k_{\bar{q}}^h,j_g;i_q^h) +  \FinPoles_X (i_q^h,j_g,k_{\bar{q}}^h) \bigg] \bigg) , \\ \nonumber
    &=& \Peup \bigg( \frac{1}{\e}  \ln \left(1 + \frac{s_{ij} s_{jk} }{s_{ik} s_{ijk}} \right)  A_3^0(i^h,j,k^h)  \bigg) , \\ \nonumber
    &=& \frac{\Af}{\e^2} \Lambda_1(i^h,j,k^h) A_3^0(i^h,j,k^h).
\end{eqnarray}
As discussed earlier, the logarithmic structure of $\Lambda_1$ suppresses all the unresolved limits present in the $A_3^0$ antenna at every order in $\e$. This structure also carries a $1/\e^2$ factor, so $\FinPoles_{\e}(i_q^h,j_g,k_{\bar{q}}^h)$ contains explicit $\e$ poles (which are important for antenna-scheme matching) but does not contribute in the unresolved limits. 
In summary,
\begin{align}
    &\PSdown_j \left( \Lambda_1(i^h,j,k^h) A_3^0(i^h,j,k^h) \right)  \to  0 \, ,\nonumber \\
    &\PCdown_{ij} \left(\Lambda_1(i^h,j,k^h) A_3^0(i^h,j,k^h) \right)  \to  0 \, ,\nonumber \\
    &\PCdown_{jk} \left(\Lambda_1(i^h,j,k^h) A_3^0(i^h,j,k^h) \right)  \to  0 \, .\nonumber
\end{align}

Finally, including the renormalisation term and combining terms together we find a compact expression for $\Arv$ given by 
\begin{equation}
\begin{split}
    \Arv(i^h,j,k^h) &= \frac{\Af}{\e^2}\bigg[ -\Gamma(1-\e)\Gamma(1+\e) \frac{S_{ij}S_{jk}}{S_{ik}} +S_{ij} G\left(\frac{s_{jk}}{s_{ik}},\e \right)  + S_{jk}G\left(\frac{s_{ij}}{s_{ik}},\e \right) \\
    & \hspace{0.5cm}  + \Lambda_1(i^h,j,k^h) \bigg] A_3^0(i^h,j,k^h) 
     - \frac{\bz}{\e} A_3^0(i^h,j,k^h)\\
    & \hspace{0.5cm} + S_{ij} \Af \frac{(s_{ijk}-\e s_{jk} )}{s_{ij}s_{ijk}} \frac{1}{2 (1-2\e)} 
    +  S_{jk} \Af \frac{(s_{ijk}-\e s_{ij} )}{s_{jk}s_{ijk}}\frac{1}{2 (1-2\e)} \, .
\end{split}
\end{equation}

We see that
\begin{align}
    \PSdown_j \Arv(i^h,j,k^h) &= S_g^{(1)}(i^h,j,k^h) \, , \\
    \PCdown_{ij} \Arv(i^h,j,k^h) &= P_{qg}^{(1)}(i^h,j) \, , \\
    \PCdown_{jk} \Arv(i^h,j,k^h) &= P_{qg}^{(1)}(k^h,j) \, ,   
\end{align}
and 
\begin{align}
    \Pedown \Arv(i^h,j,k^h) &= \Pedown T(i_q^h,j_g,k_{\bar{q}}^h)\, .
\end{align}

\section{One-loop single unresolved limits}
\label{sec:limits}

The universal soft and collinear factorisation properties of multiparticle real-virtual amplitudes have been well studied in the literature~\cite{Bern:1994zx,Bern:1998sc,Kosower:1999rx,Bern:1999ry}.
In this section, we list the unrenormalised, colour-ordered unresolved factors at one loop in conventional dimensional regularisation (CDR), which are consistent with the formulations in \cite{Kosower:1999rx,Catani:2000pi,Weinzierl:2003ra,Badger:2004uk,Gehrmann-DeRidder:2005btv,Bern:1999ry,Duhr:2013msa}. 
The overall factor $\Af$ ensures that the antennae constructed here have the same normalisation as those in Ref.~\cite{Gehrmann-DeRidder:2005svg}.

The full-colour (unrenormalised) one-loop soft factor is given by 
\begin{equation}
    \mathbf{S}_g^{(1)} (i^h,j,k^h) = N_c \Sgone(i^h,j,k^h) - \frac{1}{N_c} \Sgtone(i^h,j,k^h) + N_F \Sghone(i^h,j,k^h),
\end{equation}
where
\begin{align}
\label{eq:Sg1}
    \Sgone(i^h,j,k^h) &= - \Af\frac{\Gamma(1-\e)\Gamma(1+\e)}{\e^2} \frac{S_{ij}S_{jk}}{S_{ik}}  \Sgzero (i^h,j,k^h)  \, , \\
    \Sgtone(i^h,j,k^h) &= 0 \, , \\
    \Sghone(i^h,j,k^h) &= 0 \, ,    
\end{align}
and formally we define any soft factor, where particle $b$ with momentum $j$ is not a gluon, as zero. The tree-level single-unresolved limits are given in Appendix~\ref{app:treelimits}. 

In general, the full-colour (unrenormalised) one-loop splitting function is decomposed by
\begin{equation}
    \mathbf{P}_{ab}^{(1)}(i,j) = N_c P_{ab}^{(1)}(i,j) - \frac{1}{N_c} \widetilde{P}_{ab}^{(1)}(i,j) + N_F \widehat{P}_{ab}^{(1)}(i,j) \, .
\end{equation}
As at leading-order (see Appendix~\ref{app:treelimits}), we organise the splitting functions according to which particle is a hard-radiator.  This means that $\mathbf{P}_{ab}^{(1)}(i^h, j)$ is {\em not singular} in the limit where the hard radiator $a$ becomes soft and is directly related to the usual spin-averaged one-loop splitting functions given in terms of the momentum fraction carried by particle $j$ ($\xj$), defined with reference to the third particle in the antenna. 

The $q \to q g$ one-loop splitting functions for $i$ being the hard radiator are given by
\begin{align}
    \Pqgone(i^h,j)  &= \frac{1}{s_{ij}} \Pqgone(\xj)\, ,  \\
    \Pqgtone(i^h,j) &= \frac{1}{s_{ij}} \Pqgtone(\xj)\, ,  \\
    \Pqghone(i^h,j) &= 0 \, ,
\end{align}
with
\begin{align}
    \Pqgone(\xj) &= S_{ij} \frac{\Af}{\e^2} \bigg[ - \Gamma(1-\e) \Gamma(1+\e) \left(\frac{1-x_j}{x_j}\right)^\e + G\left(\frac{x_j}{1-x_j},\e\right) \bigg] \Pqg(\xj) \nonumber \\
    &\hspace{0.5cm} +  S_{ij} \Af\frac{(1-x_j \e )}{2(1-2\e)}
    \, , \label{eq:Pqg1} \\
    \Pqgtone(\xj) &= 
     - S_{ij} \frac{\Af}{\e^2}  G\left(\frac{x_j}{1-x_j},-\e\right) \Pqg(\xj) -  S_{ij} \Af \frac{(1-x_j \e )}{2(1-2\e)}
    \, .
\end{align}
Here $G(w,\e)$ is defined in Eq.~\eqref{eq:Ghyperdef}.  It has the property that it vanishes as $\xj \to 0$ and has the $\e$-expansion,
\begin{equation}
    G\left(\frac{x_j}{1-x_j},\e\right) = \e \ln\omxj -\e^2 \Li_2\left(\frac{-\xj}{1-\xj}\right) + \order{\e^3}.
\end{equation}

The tree-level splitting function $\Pqg$ is given in Eq.~\eqref{eq:Pqg}.
In the complementary case of $j$ being the hard radiator, all splitting functions vanish identically,
\begin{equation}
    \Pqgone(i,j^h) = 0 \, , \quad
    \Pqgtone(i,j^h) = 0 \, , \quad
    \Pqghone(i,j^h) = 0 \, . 
\end{equation}

The $g \to q \qb$ one-loop splitting functions are 
\begin{align}
    \Pqqone(i^h,j)  &= \frac{1}{s_{ij}} \Pqqone(\xj)\, , \\
    \Pqqtone(i^h,j) &= \frac{1}{s_{ij}} \Pqqtone(\xj) \, , \\
    \Pqqhone(i^h,j) &= \frac{1}{s_{ij}} \Pqqhone(\xj) \, ,
\end{align}
and
\begin{align}
    \Pqqone(i,j^h)  &= \frac{1}{s_{ij}} \Pqqone\omxj , \\ 
    \Pqqtone(i,j^h) &= \frac{1}{s_{ij}} \Pqqtone\omxj \, , \\
    \Pqqhone(i,j^h) &= \frac{1}{s_{ij}} \Pqqhone\omxj \, .
\end{align}
The one-loop splitting functions are given by
\begin{align}
    \Pqqone(\xj) &=  S_{ij} \frac{\Af}{ \e ^2} \bigg[ - \Gamma(1-\e) \Gamma(1+\e) \left( \frac{1-x_j}{x_j} \right)^\e + 1 + G\left(\frac{x_j}{1-x_j},\e\right) - G\left(\frac{x_j}{1-x_j},-\e\right) \nonumber \\
    &\hspace{1.75cm} + \frac{\e (13 - 8 \e)}{2(3-2\e)(1-2\e)} \bigg] \Pqq (\xj) \, , \\    
    \Pqqtone(\xj) &= - S_{ij} \Af \bigg[   \frac{1}{ \e^2} + \frac{3+2\e}{2 \e (1-2 \e)}   \bigg] \Pqq (\xj) \, , \\
    \Pqqhone(\xj) &=  S_{ij} \Af \bigg[ 
    - \frac{2(1-\e)}{\e (3-2 \e)(1-2\e)} \bigg] \Pqq (\xj)  \, ,
\end{align}
with $\Pqq$ defined in Eq.~\eqref{eq:Pqq}. Note that the symmetry of $\Pqqone$ is preserved:
\begin{equation}
     \Pqqone\omxj \equiv  \Pqqone(\xj).
\end{equation}

Finally, the $g \to g g$ one-loop splitting functions for $i$ being the hard radiator are given by
\begin{align}
    \Pggone(i^h,j) &= \frac{1}{s_{ij}} \Pggonesub(\xj)\, ,  \\
    \Pggtone(i^h,j) &= 0 \, , \\
    \Pgghone(i^h,j) &= \frac{1}{s_{ij}} \Pgghonesub(\xj), \, 
\end{align}
while when $j$ is the hard radiator,
\begin{align}
    \Pggone(i,j^h) &= \frac{1}{s_{ij}} \Pggonesub\omxj \, , \\
    \Pggtone(i,j^h) &= 0 \, , \\
    \Pgghone(i,j^h) &= \frac{1}{s_{ij}} \Pgghonesub\omxj \, .
\end{align}
Here, the one-loop splitting functions are given in terms of the tree-level splitting function $\PggS$ given in Eq.~\eqref{eq:PggS},
\begin{align}
    \Pggonesub(\xj) &= S_{ij} \frac{\Af}{\e^2} \bigg[ - \Gamma(1-\e) \Gamma(1+\e) \left(\frac{1-x_j}{x_j}\right)^\e + G\left(\frac{x_j}{1-x_j},\e\right) 
\nonumber \\
    &\hspace{1.5cm}    - G\left(\frac{x_j}{1-x_j},-\e\right) \bigg] \PggSzero (\xj) \nonumber \\
    &\hspace{0.5cm} + S_{ij} \Af \frac{(1-2 x_j (1-x_j) \e)}{2(1-\e)(1-2\e)(3-2\e)}
    \, , \\
    \Pgghonesub(\xj) &= S_{ij} \Af \left(\frac{-(1-2 x_j (1-x_j) \e)}{2(1-\e)^2(1-2\e)(3-2\e)}\right) \, , 
\end{align}
and satisfy
\begin{align}
   \Pggone(\xj) &= \Pggonesub(\xj) +  \Pggonesub\omxj \, , \\
   \Pgghone(\xj) &= \Pgghonesub(\xj) +  \Pgghonesub\omxj \, .
\end{align}

\section{Real-virtual antenna functions}
\label{sec:X31}

In this section, we give compact expressions for the full set of real-virtual antennae.  In deriving these antennae, we have made use of MAPLE, hypexp~\cite{Huber:2005yg,Huber:2007dx} and FORM~\cite{Vermaseren:2000nd,Kuipers:2012rf}. 

\subsection{Quark-antiquark antennae}

As shown in Table~\ref{tab:X31}, there are three one-loop three-parton antennae with quark-antiquark parents that describe the emission of a gluon, organised by colour structure: $\Arv$, $\Atrv$, and $\Ahrv$. The antenna functions constructed here are directly related to the antenna functions given in Ref.~\cite{Gehrmann-DeRidder:2005btv} by 
\begin{align}
    \Arvold (i_q,j_g,k_{\bar{q}}) &\sim \Arv (i_q^h,j_g,k_{\bar{q}}^h) \, , \\
    \Atrvold (i_q,j_g,k_{\bar{q}}) &\sim \Atrv (i_q^h,j_g,k_{\bar{q}}^h) \, , \\
    \Ahrvold (i_q,j_g,k_{\bar{q}}) &\sim \Ahrv (i_q^h,j_g,k_{\bar{q}}^h) \, ,
\end{align}
where $\sim$ means that they contain the same limits as $j_g$ becomes unresolved, although they may contain different $\e$ poles. 

In order to build these antennae using the algorithm in Section~\ref{sec:algorithm}, we identify the particles included in the antenna to specify the limits encapsulated by the $X_3^1$ and identify the target poles for the $X_3^1$ factorising on to the respective $X_3^0$ (here $A_3^0$). The resulting formula from the algorithm (copied from above) is given by
\begin{equation}
\begin{split}
    \Arv(i^h,j,k^h) &= \frac{\Af}{\e^2}\bigg[ -\Gamma(1-\e)\Gamma(1+\e) \frac{S_{ij}S_{jk}}{S_{ik}} +S_{ij} G\left(\frac{s_{jk}}{s_{ik}},\e \right)  + S_{jk}G\left(\frac{s_{ij}}{s_{ik}},\e \right) \\
    & \hspace{0.5cm}  + \Lambda_1(i^h,j,k^h) \bigg] A_3^0(i^h,j,k^h) 
    - \frac{\bz}{\e} A_3^0(i^h,j,k^h) 
    \\
    & \hspace{0.5cm} + S_{ij} \Af \frac{(s_{ijk}-\e s_{jk} )}{s_{ij}s_{ijk}} \frac{1}{2 (1-2\e)} 
    +  S_{jk} \Af \frac{(s_{ijk}-\e s_{ij} )}{s_{jk}s_{ijk}}\frac{1}{2 (1-2\e)} \, .
\end{split}
\label{eqn:A31}
\end{equation}

Integrating over the single-unresolved antenna phase space (more details are given in Section~\ref{app:integration}), we yield the integrated antenna
\begin{align}
{\cal A}_3^1 (s_{ijk}) &= S_{ijk}^2\Biggl [
-\frac{1}{4\e^4}
-\frac{31}{12\e^3}
+\frac{1}{\e^2} \left(
-\frac{53}{8}
+\frac{11}{24}\pi^2
\right)
+\frac{1}{\e} \left(
-\frac{659}{24}
+\frac{22}{9}\pi^2
+\frac{23}{3}\zeta_3
\right) \nonumber \\
&\hspace{2.5cm}
+ \left(
-\frac{1345}{12}
+\frac{199}{24}\pi^2
+\frac{635}{18}\zeta_3
+\frac{13}{1440}\pi^4
\right)
+ \order{\e}\Biggr] \, .
\label{eq:A31int}
\end{align}
This expansion differs from $\calArvold$ in Eq.~(5.18) of Ref.~\cite{Gehrmann-DeRidder:2005btv}, starting from the rational part at $\order{1/\e}$. In a similar way to the constructed $\A$ in Ref.~\cite{paper2}, this is simply because the $A_3^0$ given in Ref.~\cite{paper2} differs at $\order{\e}$ from $\Xold{A}$ of Ref.~\cite{Gehrmann-DeRidder:2005btv}. The choice of $A_3^0$ impacts the $\e$ poles of $\Arv$ at both the unintegrated and integrated levels, because $A_3^0$ factorises onto explicit $1/\e^2$ poles in Eq.~\eqref{eqn:A31}. If instead the original $\Xold{A}$, of Ref.~\cite{Gehrmann-DeRidder:2005btv}, is used in Eq.~\eqref{eqn:A31}, the integrated antenna in Eq.~\eqref{eq:A31intoldX3} contains exactly the same poles as $\calArvold$ in Eq.~(5.18) of Ref.~\cite{Gehrmann-DeRidder:2005btv} and differs only at $\order{\e^0}$. 

Similarly, for the sub-leading-colour $q \bar{q}$ antenna,
\begin{align}
    \Atrv(i^h,j,k^h) &= -\frac{\Af}{\e^2} \bigg[ S_{ij} G\left(\frac{s_{jk}}{s_{ik}},-\e \right)  + S_{jk}G\left(\frac{s_{ij}}{s_{ik}},-\e \right)  - \Lambda_1(i^h,j,k^h) \bigg] A_3^0(i^h,j,k^h) \nonumber \\
    &\hspace{1cm} - S_{ij} \Af \frac{(s_{ijk}-\e s_{jk} )}{s_{ij}s_{ijk}} \frac{1}{2 (1-2\e)} 
    - S_{jk} \Af \frac{(s_{ijk}-\e s_{ij} )}{s_{jk}s_{ijk}}\frac{1}{2 (1-2\e)} \, ,
\label{eqn:At31}
\end{align}
and after integration we find the expression,
\begin{align}
\widetilde{{\cal A}}_3^1 (s_{ijk}) &= S_{ijk}^2\Biggl [
+\frac{1}{\e^2} \left(
-\frac{5}{8}
+\frac{1}{6}\pi^2
\right)
+\frac{1}{\e} \left(
-\frac{19}{4}
+\frac{1}{4}\pi^2
+7\zeta_3
\right) \nonumber \\
&\hspace{2.5cm}
+ \left(
-\frac{447}{16}
+\frac{29}{16}\pi^2
+\frac{21}{2}\zeta_3
+\frac{7}{60}\pi^4
\right)
+ \order{\e}\Biggr] \, .
\label{eq:A31tint}
\end{align}
This expansion only differs from $\calAtrvold$ in Eq.~(5.19) of Ref.~\cite{Gehrmann-DeRidder:2005btv} at $\order{\e^0}$. In this case, the choice of $A_3^0$ does not impact the $\e$ poles of $\Atrv$ because they are at most $1/\e$ and the $A_3^0$ given in Ref.~\cite{paper2} differs only at $\order{\e}$ from $\Xold{A}$ of Ref.~\cite{Gehrmann-DeRidder:2005btv}. 

For the quark-loop $q \bar{q}$ antenna, there are no unrenormalised unresolved limits and so the antenna is simply a renormalisation term:
\begin{equation}
    \Ahrv(i^h,j,k^h) = - \frac{\bzf}{\e}   A_3^0(i^h,j,k^h) \, .
\label{eqn:Ah31}
\end{equation}
The integrated version is given by
\begin{align}
\widehat{{\cal A}}_3^1 (s_{ijk}) &= S_{ijk}^2\Biggl [
+\frac{1}{3\e^3}
+\frac{1}{2\e^2}
+\frac{1}{\e} \left(
\frac{19}{12}
-\frac{7}{36}\pi^2
\right) \nonumber \\
&\hspace{2.5cm}
+ \left(
\frac{113}{24}
-\frac{7}{24}\pi^2
-\frac{25}{9}\zeta_3
\right)
+ \order{\e}\Biggr] \, ,
\label{eq:A31hint}
\end{align}
which only differs from $\calAhrvold$ in Eq.~(5.20) of Ref.~\cite{Gehrmann-DeRidder:2005btv} at $\order{\e^0}$. In this case, the choice of $A_3^0$ does not impact the $\e$ poles of $\Ahrv$ because they are at most $1/\e$ and the $A_3^0$ given in Ref.~\cite{paper2} differs only at $\order{\e}$ from $\Xold{A}$ of Ref.~\cite{Gehrmann-DeRidder:2005btv}.

\subsection{Quark-gluon antennae}

As shown in Table~\ref{tab:X31}, there are six one-loop three-parton antennae with quark-gluon parents organised by colour structure: $\Drv$, $\Dtrv$, $\Dhrv$, $\Erv$, $\Etrv$, and $\Ehrv$. The antenna functions constructed here are directly related to the antenna functions given in Ref.~\cite{Gehrmann-DeRidder:2005btv} by 
\begin{align}
    \Drvold (i_q,j_g,k_{g}) &\sim \Drv(i_q^h,j_g,k_{g}^h)
    +\Dtrv (i_q^h,j_g,k_{g}^h) + (j \leftrightarrow k) \, , \label{eq:Doldnew} \\
    \Dhrvold (i_q,j_g,k_{\bar{q}}) &\sim \Dhrv (i_q^h,j_g,k_{g}^h) + \Dhrv (i_q^h,k_g,j_{g}^h) \, , \\
    \Ervold (i_q,j_{\bar{Q}},k_{Q}) &\sim \Erv(i_q^h,j_{\bar{Q}},k_{Q}^h) \, , \\
    \Etrvold (i_q,j_{\bar{Q}},k_{Q}) &\sim \Etrv(i_q^h,j_{\bar{Q}},k_{Q}^h) \, , \\ 
    \Ehrvold (i_q,j_{\bar{Q}},k_{Q}) &\sim \Ehrv(i_q^h,j_{\bar{Q}},k_{Q}^h) \, .
\end{align}
Note that $\Drvold$ was extracted from an effective Lagrangian describing heavy neutralino decay into a gluino-gluon pair, where the gluino plays the role of the quark~\cite{Gehrmann-DeRidder:2005svg}. Firstly, $\Drvold$ contains unresolved configurations where either of the gluons can be soft, so this is decomposed here such that only one gluon can be soft. Secondly, the extracted antennae $\Drvold$ (and $\Dold$) contain both leading-colour and sub-leading colour limits and they receive special treatment in the antenna scheme. In Ref.~\cite{paper2}, we effectively split $\Dold$ into a combination of $\D$ and $\Dt$ antennae and we perform a similar decomposition here of $\Drvold$ into $\Drv$ and $\Dtrv$. Due to the absence of a sub-leading colour $\Dtrvold$ antenna, we only have target poles, $T(i_q^h,j_g,k_g^h)$, for the combination of $\Drv(i_q^h,j_g,k_g^h)+\Dtrv(i_q^h,j_g,k_g^h)$. We choose to place the resulting $\FinPoles_{\e}$ term in the formula for $\Drv$. To recap, the combination of $\Drv(i_q^h,j_g,k_g^h)+\Dtrv(i_q^h,j_g,k_g^h)$ have been used to match $\e$ poles in the existing antenna-subtraction-scheme, while $\Drv(i_q^h,j_g,k_g^h)$ contains the leading-colour limits when $j$ is unresolved and $\Dtrv(i_q^h,j_g,k_g^h)$ contains the sub-leading-colour limits when $j$ is unresolved. This means the two antennae $\Drv,\Dtrv$ could in principle be used independently in subtraction terms to cancel relevant one-loop unresolved limits, but the $\e$-pole cancellation may require specific attention. 

%
 %
%
%
%
  %

%
%
%

The $\Drv$ formula is given by 
\begin{eqnarray}
\label{eqn:D31}
    \Drv(i^h,j,k^h) &=& \frac{\Af}{\e^2}\bigg[ -\Gamma(1-\e)\Gamma(1+\e) \frac{S_{ij}S_{jk}}{S_{ik}} \\ \nonumber
    &&+S_{ij} G\left(\frac{s_{jk}}{s_{ik}},\e \right) +S_{jk} G\left(\frac{s_{ij}}{s_{ik}},\e \right)  - S_{jk}G\left(\frac{s_{ij}}{s_{ik}},-\e \right) \\ \nonumber
    && + 2 \Lambda_1(i^h,j,k^h) \bigg] D_3^0 (i^h,j,k^h) 
    - \frac{\bz}{\e} D_3^0 (i^h,j,k^h)
    \\ \nonumber
    && + S_{ij} \Af \frac{(s_{ijk}-\e s_{jk} )}{s_{ij}s_{ijk}} \frac{1}{2 (1-2\e)}  + \frac{\Af}{2(1-\e)(1-2\e)(3-2\e)} \frac{S_{jk}}{s_{jk}} \left( 1-2 \e \frac{s_{ij}s_{ik}}{s_{ijk}^2} \right)  ,
\end{eqnarray}
and after integration we find the expression
\begin{eqnarray}
\label{eq:d31int}
{\cal D}_3^1 (s_{ijk}) &=& S_{ijk}^2\Biggl [
-\frac{1}{4\e^4}
-\frac{8}{3\e^3}
+\frac{1}{\e^2} \left(
-\frac{1109}{144}
+\frac{13}{24}\pi^2
\right)
+\frac{1}{\e} \left(
-\frac{14603}{432}
+\frac{49}{18}\pi^2
+\frac{73}{6}\zeta_3
\right)\nonumber \\
&& \hspace{2cm}
 + \left(
-\frac{7985}{54}
+\frac{8561}{864}\pi^2
+\frac{535}{12}\zeta_3
+\frac{79}{480}\pi^4
\right)
 + \order{\e}\Biggr].
\end{eqnarray}
	 %
The $\Dtrv$ formula is given by 
\begin{eqnarray}
\label{eqn:Dt31}
    \Dtrv(i^h,j,k^h) &=& -\frac{\Af}{\e^2} S_{ij} G\left(\frac{s_{jk}}{s_{ik}},-\e \right) D_3^0(i^h,j,k^h) \\ \nonumber
    && - S_{ij} \Af \frac{(s_{ijk}-\e s_{jk} )}{s_{ij}s_{ijk}} \frac{1}{2 (1-2\e)}  ,
\end{eqnarray}
and after integration we find the expression
\begin{eqnarray}
\label{eq:d31tint}
\widetilde{{\cal D}}_3^1 (s_{ijk}) &=& S_{ijk}^2\Biggl [
+\frac{1}{\e^2} \left(
-\frac{5}{16}
+\frac{1}{12}\pi^2
\right)
+\frac{1}{\e} \left(
-\frac{77}{48}
+\frac{11}{72}\pi^2
+\frac{5}{2}\zeta_3
\right)\nonumber \\
&& \hspace{2cm}
 + \left(
-\frac{983}{144}
+\frac{941}{864}\pi^2
+\frac{55}{12}\zeta_3
-\frac{7}{180}\pi^4
\right)
 + \order{\e}\Biggr].
\end{eqnarray}
The combination of $2(\calDrv + \calDtrv)$ differs from $\calDrvold$ in Eq.~(6.22) of Ref.~\cite{Gehrmann-DeRidder:2005btv}, starting from $\order{1/\e^2}$. In a similar way to the constructed $\D$ and $\Dt$ in Ref.~\cite{paper2}, this is simply because the $D_3^0$ given in Ref.~\cite{paper2} differs at $\order{\e^0}$ from $\Xold{d}$ of Ref.~\cite{Gehrmann-DeRidder:2005btv}. The choice of $D_3^0$ impacts the $\e$ poles of $\Drv$ and $\Dtrv$ at both the unintegrated and integrated levels because $D_3^0$ factorises onto explicit $1/\e^2$ poles in Eq.~\eqref{eqn:D31} and Eq.~\eqref{eqn:Dt31}. 
          
The quark-loop $qg$ antenna function is given by
\begin{align}
    \Dhrv(i^h,j,k^h) &= -\Af \frac{S_{jk}}{s_{jk}} \frac{1}{2(1-\e)^2(1-2\e)(3-2\e)} \left(1-2 \e \frac{s_{ij}s_{ik}}{s_{ijk}^2} \right) \nonumber \\
    &\hspace{0.5cm} - \frac{\bzf}{\e}    D_3^0(i^h,j,k^h),
\label{eqn:Dh31}
\end{align}
and after integration we find the expression
\begin{align}
\widehat{{\cal D}}_3^1 (s_{ijk}) &= S_{ijk}^2\Biggl [
+\frac{1}{3\e^3}
+\frac{5}{9\e^2}
+\frac{1}{\e} \left(
\frac{125}{72}
-\frac{7}{36}\pi^2
\right)  + \left(
\frac{97}{18}
-\frac{35}{108}\pi^2
-\frac{25}{9}\zeta_3
\right) 
+ \order{\e}\Biggr] \, .
\label{eq:d31hint}
\end{align}
This expansion differs from $\calDhrvold/2$ in Eq.~(6.23) of Ref.~\cite{Gehrmann-DeRidder:2005btv}, starting from the rational part at $\order{1/\e}$. In the $\Dhrv$ formula, the poles are at most $1/\e$ and they only appear in the renormalisation term. The finite difference between $D_3^0$ and $\Xold{d}$ from Ref.~\cite{Gehrmann-DeRidder:2005btv} therefore only impacts the $\order{1/\e}$ poles. 

%

The $\Erv$-type antennae contain contributions to only one limit -- the $jk$ collinear limit, when the $\bar{Q}Q$ pair become collinear and as such they are simpler expressions than the others. The first antenna is given by
\begin{align}
    \Erv(i^h,j,k^h) &= - \frac{\Af}{\e^2} \bigg[ S_{jk} G\left(\frac{s_{ij}}{s_{ik}},-\e \right) + S_{jk} G\left(\frac{s_{ik}}{s_{ij}},-\e \right) \nonumber \\
    &\hspace{0.5cm} - S_{jk} \frac{\e (13 - 8 \e)}{2(3-2\e)(1-2\e)} - 2 \Lambda_2(i^h,j,k^h) \bigg] E_3^0 (i^h,j,k^h) 
    - \frac{\bz}{\e} E_3^0 (i^h,j,k^h)
    \, , 
\label{eqn:E31}
\end{align}
where the $\Lambda_2/\e^2$ term suppresses the only limit in the $E_3^0$ to which it factorises (the $jk$ collinear limit) and thus only affects the $\e$ pole structure of $\Erv$. 
After integration we find the expression
\begin{align}
{\cal E}_3^1 (s_{ijk}) &= S_{ijk}^2\Biggl [
+\frac{11}{18\e^2}
+\frac{1}{\e} \left(
\frac{56}{27}
-\frac{1}{9}\pi^2
\right) + \left(
\frac{4111}{432}
-\frac{131}{216}\pi^2
-4\zeta_3
\right)
+ \order{\e}\Biggr] \, ,
\label{eq:E31int}
\end{align}
which differs from $\calErvold$ in Eq.~(6.34) of Ref.~\cite{Gehrmann-DeRidder:2005btv}, starting from the rational part at $\order{1/\e}$. In the $\Erv$ formula, the poles are at most $1/\e$. The finite difference between $E_3^0$ and $\Xold{E}$ from Ref.~\cite{Gehrmann-DeRidder:2005btv} therefore only impacts the $\order{1/\e}$ poles. 

%

The sub-leading-colour antenna is given by
\begin{equation}
    \Etrv(i^h,j,k^h) = -\Af S_{jk} \bigg[ \frac{1}{\e^2} + \frac{(3+2\e)}{2 \e (1-2 \e)} \bigg] E_3^0 (i^h,j,k^h) \, , 
\label{eqn:Et31}
\end{equation}
and after integration we find the expression
\begin{align}
\widetilde{{\cal E}}_3^1 (s_{ijk}) &= S_{ijk}^2\Biggl [
+\frac{1}{6\e^3}
+\frac{13}{18\e^2}
+\frac{1}{\e} \left(
\frac{613}{216}
-\frac{1}{4}\pi^2
\right) + \left(
\frac{3359}{324}
-\frac{13}{12}\pi^2
-\frac{31}{9}\zeta_3
\right)
+ \order{\e}\Biggr],
\label{eq:E31tint}
\end{align}
which differs from $\calEtrvold$ in Eq.~(6.35) of Ref.~\cite{Gehrmann-DeRidder:2005btv}, starting from the rational part at $\order{1/\e^2}$. In the $\Etrv$ formula, the poles are at most $1/\e^2$. The finite difference between $E_3^0$ and $\Xold{E}$ from Ref.~\cite{Gehrmann-DeRidder:2005btv} therefore impacts the $\order{1/\e^2}$ poles. 

The quark-loop antenna is given by
\begin{eqnarray}
\label{eqn:Eh31}
    \Ehrv(i^h,j,k^h) &=& - \Af \bigg[ S_{jk} \frac{2(1-\e)}{\e (3-2 \e)(1-2\e)} \bigg] E_3^0 (i^h,j,k^h)  
    - \frac{\bzf}{\e}  E_3^0 (i^h,j,k^h),
\end{eqnarray}
and after integration we find the expression
\begin{eqnarray}
\label{eq:E31hint}
\widehat{{\cal E}}_3^1 (s_{ijk}) &=& S_{ijk}^2\Biggl [
+\frac{1}{4\e}
+ \left(
\frac{791}{648}
-\frac{11}{108}\pi^2
\right)
 + \order{\e}\Biggr],
\end{eqnarray}
which differs from $\calEhrvold$ in Eq.~(6.36) of Ref.~\cite{Gehrmann-DeRidder:2005btv}, starting from the rational part at $\order{1/\e}$. In the $\Ehrv$ formula, the poles are at most $1/\e$. The finite difference between $E_3^0$ and $\Xold{E}$ from Ref.~\cite{Gehrmann-DeRidder:2005btv} therefore impacts the $\order{1/\e}$ poles of $\calEhrv$, although these are the deepest poles in this case. This is because of cancellations at $\order{1/\e^2}$ between the integrals of the first and second terms in Eq.~\eqref{eqn:Eh31}. 

\subsection{Gluon-gluon antennae}

As shown in Table~\ref{tab:X31}, there are five one-loop three-parton antennae with gluon-gluon parents organised by colour structure: $\Frv$, $\Fhrv$, $\Grv$, $\Gtrv$, and $\Ghrv$. The antenna functions constructed here are directly related to the antenna functions given in Ref.~\cite{Gehrmann-DeRidder:2005btv} by 
\begin{align}
    \Frvold (i_g,j_g,k_{g}) &\sim \Frv (i_g^h,j_g,k_{g}^h) + \Frv (j_g^h,k_g,i_{g}^h) + \Frv (k_g^h,i_g,j_{g}^h) \, , \label{eq:Foldnew} \\
    \Fhrvold (i_g,j_g,k_{g}) &\sim \Fhrv (i_g^h,j_g,k_{g}^h) + \Fhrv (j_g^h,k_g,i_{g}^h) + \Fhrv (k_g^h,i_g,j_{g}^h) \, , \\ 
    \Grvold (i_g,j_{\bar{Q}},k_{Q}) &\sim \Grv (i_g^h,j_{\bar{Q}},k_{Q}^h) \, , \\
    \Gtrvold (i_g,j_{\bar{Q}},k_{Q}) &\sim \Gtrv (i_g^h,j_{\bar{Q}},k_{Q}^h) \, , \\
    \Ghrvold (i_g,j_{\bar{Q}},k_{Q}) &\sim \Ghrv (i_g^h,j_{\bar{Q}},k_{Q}^h) \, .
\end{align}
Note that $\Frvold$ was extracted from an effective Lagrangian describing Higgs boson decay into gluons~\cite{Gehrmann-DeRidder:2005alt}. This means that $\Frvold$ contains unresolved configurations where any one of the three gluons can be soft, so this is decomposed here such that only one gluon can be soft. The same discussion can be applied to $\Fhrv$. 

%
%

The resulting formula for the three-gluon one-loop antenna function at leading-colour is given by
\begin{align}
    \Frv(i^h,j,k^h) &= \frac{\Af}{\e^2}\bigg[ -\Gamma(1-\e)\Gamma(1+\e) \frac{S_{ij}S_{jk}}{S_{ik}} \nonumber \\
    &\hspace{1cm} + S_{ij} G\left(\frac{s_{jk}}{s_{ik}},\e \right) - S_{ij}G\left(\frac{s_{jk}}{s_{ik}},-\e \right) \nonumber \\
    &\hspace{1cm} + S_{jk} G\left(\frac{s_{ij}}{s_{ik}},\e \right) - S_{jk}G\left(\frac{s_{ij}}{s_{ik}},-\e \right) \label{eqn:F31} \\
    &\hspace{1cm} + 2 \Lambda_1(i^h,j,k^h) \bigg] F_3^0 (i^h,j,k^h) 
    - \frac{\bz}{\e} F_3^0 (i^h,j,k^h)
    \nonumber \\
    &\hspace{0.5cm} + \frac{\Af}{2(1-\e)(1-2\e)(3-2\e)} \bigg[ \frac{S_{ij}}{s_{ij}} \left(1-2 \e \frac{s_{jk}s_{ik}}{s_{ijk}^2} \right) + \frac{S_{jk}}{s_{jk}} \left( 1-2 \e \frac{s_{ij}s_{ik}}{s_{ijk}^2} \right) \bigg] \, , \nonumber
\end{align}
and after integration we find the expression
\begin{align}
{\cal F}_3^1 (s_{ijk}) &= S_{ijk}^2\Biggl [
-\frac{1}{4\e^4}
-\frac{11}{4\e^3}
+\frac{1}{\e^2} \left(
-\frac{79}{9}
+\frac{5}{8}\pi^2
\right)
+\frac{1}{\e} \left(
-\frac{8339}{216}
+\frac{55}{18}\pi^2
+\frac{44}{3}\zeta_3
\right) \nonumber \\
&\hspace{2.5cm}
+ \left(
-\frac{73169}{432}
+\frac{5137}{432}\pi^2
+\frac{473}{9}\zeta_3
+\frac{181}{1440}\pi^4
\right)
+ \order{\e}\Biggr] \, .
\label{eq:f31int}
\end{align}
This expansion differs from $\calFrvold/3$ in Eq.~(7.22) of Ref.~\cite{Gehrmann-DeRidder:2005btv}, starting from the rational part at $\order{1/\e^2}$. In the $\Frv$ formula, the poles are at most $1/\e^2$. The finite difference between $F_3^0$ and $\Xold{f}$ from Ref.~\cite{Gehrmann-DeRidder:2005btv} therefore impacts the $\order{1/\e^2}$ poles. 


The quark-loop antenna function is given by
\begin{align}
    \Fhrv(i^h,j,k^h) &= \frac{\Af}{2(1-\e)^2(1-2\e)(3-2\e)} \bigg[ \frac{S_{ij}}{s_{ij}} \left( 1-2 \e \frac{s_{jk}s_{ik}}{s_{ijk}^2} \right) \nonumber \\
    &\hspace{0.5cm} + \frac{S_{jk}}{s_{jk}} \left( 1-2 \e \frac{s_{ij}s_{ik}}{s_{ijk}^2} \right) \bigg] - \frac{\bzf}{\e}  F_3^0(i^h,j,k^h) \, ,
\label{eqn:Fh31}
\end{align}
and after integration we find the expression
\begin{align}
\widehat{{\cal F}}_3^1 (s_{ijk}) &= S_{ijk}^2\Biggl [
+\frac{1}{3\e^3}
+\frac{11}{18\e^2}
+\frac{1}{\e} \left(
\frac{17}{9}
-\frac{7}{36}\pi^2
\right) 
+ \left(
\frac{437}{72}
-\frac{77}{216}\pi^2
-\frac{25}{9}\zeta_3
\right)
+ \order{\e}\Biggr] \, .
\label{eq:f31hint}
\end{align}
This expansion differs from $\calFhrvold/3$ in Eq.~(7.23) of Ref.~\cite{Gehrmann-DeRidder:2005btv}, starting from the rational part at $\order{1/\e}$. In the $\Fhrv$ formula, the poles are at most $1/\e$ and they only appear in the renormalisation term. The finite difference between $F_3^0$ and $\Xold{f}$ from Ref.~\cite{Gehrmann-DeRidder:2005btv} therefore only impacts the $\order{1/\e}$ poles. 

%

The formula for the one-loop gluon-splitting $gg$ antenna function at leading-colour is given by
\begin{align}
    \Grv(i^h,j,k^h) &= - \frac{\Af}{\e^2} \bigg[ S_{jk} G\left(\frac{s_{ij}}{s_{ik}},-\e \right) + S_{jk} G\left(\frac{s_{ik}}{s_{ij}},-\e \right) \label{eqn:G31} \nonumber  \\
    &\hspace{1.3cm} - S_{jk} \frac{\e (13 - 8 \e)}{2(3-2\e)(1-2\e)} - 2 \Lambda_2(i^h,j,k^h) \bigg] G_3^0 (i^h,j,k^h) 
    - \frac{\bz}{\e} G_3^0 (i^h,j,k^h)\, , \nonumber \\
\end{align}
and after integration we find the expression
\begin{equation}
\label{eq:G31int}
{\cal G}_3^1 (s_{ijk}) = S_{ijk}^2\Biggl [
+\frac{11}{18\e^2}
+\frac{1}{\e} \left(
\frac{56}{27}
-\frac{1}{9}\pi^2
\right)
+ \left(
\frac{4111}{432}
-\frac{131}{216}\pi^2
-4\zeta_3
\right)
 + \order{\e}\Biggr].
\end{equation}
Firstly, given that $E_3^0 = G_3^0$ (from Ref.~\cite{paper2}) and that $\Erv$ and $\Grv$ encapsulate the same limits, these formulae (unintegrated and integrated) are identical for the $\Erv$- and $\Grv$- type antennae:
\begin{align}
    \Grv(i^h,j,k^h) &= \Erv(i^h,j,k^h) \, , \\ 
   \Gtrv(i^h,j,k^h) &= \Etrv(i^h,j,k^h) \, , \\ 
   \Ghrv(i^h,j,k^h) &= \Ehrv(i^h,j,k^h) \, .
\end{align}
Therefore the discussion for the $\Grv$-type antennae is the same as below Eq.~\eqref{eq:E31int}, Eq.~\eqref{eq:E31tint}, and Eq.~\eqref{eq:E31hint}, respectively. When the  $X_3^{0,\text{OLD}}$ from Ref.~\cite{Gehrmann-DeRidder:2005btv} are used, the $\Grv$- and $\Erv$- type antennae have a different pole structure but the same collinear limits. 

%

The sub-leading-colour antenna function is given by
\begin{equation}
    \Gtrv(i^h,j,k^h) = -\Af S_{jk} \bigg[ \frac{1}{\e^2} + \frac{(3+2\e)}{2 \e (1-2 \e)} \bigg] G_3^0 (i^h,j,k^h) \, ,
\label{eqn:Gt31}
\end{equation}
and after integration we find the expression
\begin{align}
\widetilde{{\cal G}}_3^1 (s_{ijk}) &= S_{ijk}^2\Biggl [
+\frac{1}{6\e^3}
+\frac{13}{18\e^2}
+\frac{1}{\e} \left(
\frac{613}{216}
-\frac{1}{4}\pi^2
\right)
+ \left(
\frac{3359}{324}
-\frac{13}{12}\pi^2
-\frac{31}{9}\zeta_3
\right)
+ \order{\e}\Biggr]  \, .
\label{eq:G31tint}
\end{align}
See the discussion for Eq.~\eqref{eq:E31tint}, which also applies to Eq.~\eqref{eq:G31tint}. 


The quark-loop antenna function is given by
\begin{equation}
    \Ghrv(i^h,j,k^h) = - \Af \bigg[ S_{jk} \frac{2(1-\e)}{\e (3-2 \e)(1-2\e)} \bigg] G_3^0 (i^h,j,k^h)
    - \frac{\bzf}{\e}  G_3^0 (i^h,j,k^h) \, ,
\label{eqn:Gh31}
\end{equation}
and after integration we find the expression
\begin{equation}
\widehat{{\cal G}}_3^1 (s_{ijk}) = S_{ijk}^2\Biggl [
+\frac{1}{4\e}
+ \left(
\frac{791}{648}
-\frac{11}{108}\pi^2
\right)
+ \order{\e}\Biggr] \, .
\label{eq:G31hint}
\end{equation}
See the discussion for Eq.~\eqref{eq:E31hint}, which also applies to Eq.~\eqref{eq:G31hint}. 

\section{Antenna-subtraction scheme consistency checks}
\label{sec:J22}

In the antenna subtraction scheme, the virtual (NLO) and double-virtual (NNLO) subtraction terms can be written in terms of integrated dipoles denoted by $\J{1}$ and $\J{2}$ respectively~\cite{Currie:2013vh}.  These integrated dipoles are formed by systematically combining integrated antenna-function contributions from the real and real-virtual layers (together with appropriate mass factorisation terms). The NNLO integrated dipole $\J{2}$ naturally emerges from the groups of integrated antenna functions (and mass factorisation kernels) and, together with combinations of $\J{1}$, reproduces and properly subtracts the explicit poles of the double-virtual contribution to the NNLO cross section. The integrated dipoles are therefore intimately related to Catani’s IR singularity operators~\cite{Catani:1998nv} which describe the singularities of virtual matrix elements.  It is a non-trivial check of an antenna scheme constructed directly from unresolved limits that the integrated dipoles cancel the explicit poles of the double-virtual contribution. In this section, we write down expressions for $\J{2}$ (and $\J{1}$) and show that they produce the correct pole structure.

We start from the expressions for the integrated dipoles in colour space~\cite{Chen:2022ktf},
\begin{eqnarray}
\Jcol{\ell} (\e) &=& \sum_{(i,j)}{\mathcal{J}}^{(\ell)}_2 (i,j) \textbf{T}_i\cdot\textbf{T}_j , \\ \nonumber
\Jcolb{2} (\e) &=&\sum_{(q,\bar{q})}\sum_{g}\overline{\mathcal{J}}^{(2)}_2\left(q,\bar{q}\right)\,\left(\textbf{T}_q+\textbf{T}_{\bar{q}}\right)\cdot\textbf{T}_g,
\end{eqnarray}
and further divide the ${\mathcal{J}}^{(\ell)}_2 (i,j)$ according to their colour-structures,
\begin{eqnarray}
\mathcal{J}^{(1)}_2\left(q,\bar{q}\right)&=&N_c\,\J{1}\left(q,\bar{q}\right), \\ \nonumber
\mathcal{J}^{(1)}_2\left(q,g\right)&=&N_c\J{1}\left(q,g\right)+N_F\,\Jh{1}\left(q,g\right), \\ \nonumber
\mathcal{J}^{(1)}_2\left(g,g\right)&=&N_c\J{1}\left(g,g\right)+N_F\,\Jh{1}\left(g,g\right), \\ \nonumber
\end{eqnarray}
and
\begin{eqnarray}
\mathcal{J}^{(2)}_2\left(q,\bar{q}\right)&=&N_c^2\,\J{2}\left(q,\bar{q}\right)-\Jt{2}\left(q,\bar{q}\right)+N_c N_F\Jh{2}\left(q,\bar{q}\right), \\ \nonumber
\mathcal{J}^{(2)}_2\left(q,g\right)&=&N_c^2\,\J{2}\left(q,g\right)+N_c N_F\,\Jh{2}\left(q,g\right)-\dfrac{N_F}{N_c}\Jht{2}\left(q,g\right)+N_F^2\,\Jhh{2}\left(q,g\right), \\ \nonumber
\mathcal{J}^{(2)}_2\left(g,g\right)&=&N_c^2\,\J{2}\left(g,g\right)+N_c N_F\,\Jh{2}\left(g,g\right)-\dfrac{N_F}{N_c}\Jht{2}\left(g,g\right)+N_F^2\,\Jhh{2}\left(g,g\right), \\ \nonumber
\overline{\mathcal{J}}^{(2)}_2\left(q,\bar{q}\right)&=&\dfrac{N_c^2}{2}\,\Jb{2}\left(q,\bar{q}\right)-\dfrac{1}{2}\,\Jt{2}\left(q,\bar{q}\right) .
\end{eqnarray}

In order to cancel the singularities of one- and two-loop matrix elements, 
$\Jtot{1}$ and $\Jtot{2}$ ($\Jtotb{2}$) must be related to
$\JtotOLD{1}$ and $\JtotOLD{2}$ ($\JtotbOLD{2}$) given in Ref.~\cite{Chen:2022clm}.
In particular, they must satisfy the following identities~\cite{Chen:2022ktf} which ensure that they match the known singularity structures at one and two loops.
At NLO,
\begin{equation}
\label{eq:J21I1}
    \Jtot{1} (i,j) = \JtotOLD{1} (i,j)  + \order{\e^0} ,
\end{equation}
and at NNLO,    
\begin{align}
\label{eq:J22I2}
    &\Jtot{2}\left(q,\bar{q}\right) - \frac{\Bz}{\e} \Jtot{1}\left(q,\bar{q}\right) = 
    \JtotOLD{2}\left(q,\bar{q}\right) - \frac{\Bz}{\e} \JtotOLD{1}\left(q,\bar{q}\right) + \mathcal{O}\left(\epsilon^0\right) \, , \\ 
    &\Jtot{2}\left(g,g\right) - \frac{\Bz}{\e}\Jtot{1}\left(g,g\right) =
    \JtotOLD{2}\left(g,g\right) - \frac{\Bz}{\e}\JtotOLD{1}\left(g,g\right)
    +\mathcal{O}\left(\epsilon^0\right) \, , \\
    &\Jtot{2}\left(q,g\right) + \Jtot{2}\left(g,\bar{q}\right) - 2\Jtotb{2}\left(q,\bar{q}\right) - \frac{\Bz}{\e}\left(\Jtot{1}\left(q,g\right) + \Jtot{1}\left(g,\bar{q}\right)\right) \nonumber\\
    &\hspace{4.275cm} = \JtotOLD{2}\left(q,g\right) + \JtotOLD{2}\left(g,\bar{q}\right) - 2\JtotbOLD{2}\left(q,\bar{q}\right) \\
    &\hspace{5cm} - \frac{\Bz}{\e}\left(\JtotOLD{1}\left(q,g\right) + \JtotOLD{1}\left(g,\bar{q}\right)\right)
    + \mathcal{O}\left(\epsilon^0\right) \, . \nonumber
\end{align}

Here we give the new definitions for the $\J{1}$ and $\J{2}$ pertaining to final-final configurations, which satisfy the above identities and are constructed from the integrated versions of $X_3^0$ and $X_4^0$ presented in Ref.~\cite{paper2} and the integrated $X_3^1$ constructed in this paper, thus completing the set of antenna functions required for the complete final-final (FF) NNLO subtraction scheme. The $\J{1}$ are defined by
\begin{eqnarray}
\label{eq:J21QQ}
\J{1} \left(1_{q},2_{\bar{q}}\right) &=& \cala(s_{12}), \\
\label{eq:J21QG}
\J{1}\left(1_{q},2_{g}\right) &=& \cald(s_{12}), \\
\label{eq:J21hQG}
\Jh{1}\left(1_{q},2_{g}\right) &=& \frac{1}{2}\cale(s_{12}), \\
\label{eq:J21GG}
\J{1}\left(1_{g},2_{g}\right) &=& \calf(s_{12}), \\
\label{eq:J21hGG}
\Jh{1}\left(1_{g},2_{g}\right) &=& \calg(s_{12}) ,
\end{eqnarray}
while the $\J{2}$ are defined below. 
For $q\bar{q}$ antennae, the integrated dipoles read
\begin{align}
\J{2} \left(1_{q},2_{\bar{q}}\right) &=
\calA(s_{12}) + \calArv(s_{12}) + \frac{\bz}{\e} \left(\frac{s_{12}}{\mu^2}\right)^{-\e} \cala(s_{12}) - \frac{1}{2} \left[ \cala \otimes \cala \right](s_{12}) \, ,
\label{eq:J22QQ} \\
\Jt{2} \left(1_{q},2_{\bar{q}}\right) &=
\frac{1}{2} \calAt(s_{12}) + 2\calC(s_{12}) + \calAtrv(s_{12}) - \frac{1}{2}\left[ \cala \otimes \cala \right](s_{12}) \, , \label{eq:J22tQQ} \\
\Jh{2} \left(1_{q},2_{\bar{q}}\right) &=
\calB(s_{12}) + \calAhrv(s_{12})+ \frac{\bzf}{\e} \left(\frac{s_{12}}{\mu^2}\right)^{-\e} \cala(s_{12}) \, , \label{eq:J22hQQ} \\
\Jb{2} \left(1_{q},2_{\bar{q}}\right) &=
\frac{1}{2}\calAt(s_{12}) + \calAtrv(s_{12}) - \frac{1}{2}\left[ \cala \otimes \cala \right] (s_{12})\, . \label{eq:J22barQQ}
\end{align}
For $qg$ antennae, the integrated dipoles are given by
\begin{align}
\J{2} \left(1_{q},2_{g}\right) &=
\calD(s_{12}) + \frac{1}{2} \calDt(s_{12}) + \calDrv(s_{12}) +\calDtrv(s_{12}) + \frac{\bz}{\e} \left(\frac{s_{12}}{\mu^2}\right)^{-\e} \cald (s_{12}) 
\nonumber \\
&\hspace{0.5cm}
- \left[ \cald \otimes \cald \right](s_{12}) \, , \label{eq:J22QG} \\
\Jh{2} \left(1_{q},2_{g}\right) &=
\calEa (s_{12})+ \calEb(s_{12}) + \calDhrv(s_{12}) + \frac{1}{2}\calErv (s_{12})
+ \frac{\bzf}{\e} \left(\frac{s_{12}}{\mu^2}\right)^{-\e} \cald (s_{12})\nonumber \\
&\hspace{0.5cm}
+ \frac{1}{2}\frac{\bz}{\e} \left(\frac{s_{12}}{\mu^2}\right)^{-\e} \cale(s_{12}) - \left[ \cald \otimes \cale \right](s_{12}) \, , \label{eq:J22hQG} \\
\Jhh{2}\left(1_{q},2_{g}\right) &= 
\frac{1}{2}\calEhrv(s_{12}) + 
\frac{1}{2} \frac{\bzf}{\e} \left(\frac{s_{12}}{\mu^2}\right)^{-\e} \cale(s_{12}) - \frac{1}{4}\left[ \cale \otimes \cale \right](s_{12}) \, , \label{eq:J22hhQG} \\
\Jht{2} \left(1_{q},2_{g}\right) &=
\frac{1}{2} \calEt(s_{12}) + \frac{1}{2} \calEtrv (s_{12})\, . \label{eq:J22htQG}
\end{align}
Finally, for $gg$ antennae, the integrated dipoles are
\begin{align}
\J{2} \left(1_{g},2_{g}\right) &=
\calF(s_{12}) + \frac{1}{2} \calFt(s_{12}) + \calFrv(s_{12}) + \frac{\bz}{\e} \left(\frac{s_{12}}{\mu^2}\right)^{-\e} \calf(s_{12}) \nonumber \\
&\hspace{0.5cm}- \left[ \calf \otimes \calf \right] (s_{12})\, , \label{eq:J22GG} \\
\Jh{2} \left(1_{g},2_{g}\right) &=
\calGa(s_{12}) + 2\calGb(s_{12}) + \calFhrv(s_{12}) + \calGrv(s_{12}) 
+ \frac{\bzf}{\e} \left(\frac{s_{12}}{\mu^2}\right)^{-\e} \calf(s_{12}) \nonumber \\
&\hspace{0.5cm}
+ \frac{\bz}{\e} \left(\frac{s_{12}}{\mu^2}\right)^{-\e} \calg
(s_{12})
- 2\left[ \calf \otimes \calg \right] (s_{12})\, , \label{eq:J22hGG} \\
\Jhh{2} \left(1_{g},2_{g}\right) &=
\frac{1}{2}\calH (s_{12})+ \calGhrv (s_{12})+
 \frac{\bzf}{\e} \left(\frac{s_{12}}{\mu^2}\right)^{-\e} \calg (s_{12}) - \left[ \calg \otimes \calg \right] (s_{12})\, , \label{eq:J22hhGG} \\
\Jht{2}\left(1_{g},2_{g}\right) &=
\calGt(s_{12}) + \calGtrv (s_{12})\, . \label{eq:J22htGG}
\end{align}
Note that, apart from certain well understood rescalings (such as $(1/3){\cal F}_3^0 \mapsto {\cal F}_3^0$, $(1/2){\cal D}_3^0 \mapsto {\cal D}_3^0$ and so on), Eqs.~\eqref{eq:J21QQ}--\eqref{eq:J22htGG} have a very similar structure to those appearing in Ref.~\cite{Currie:2013vh}. 

We observe that Eqs.~\eqref{eq:J21QQ}--\eqref{eq:J21hGG} satisfy Eq.~\eqref{eq:J21I1}.  This is no surprise since the integrated single-real antenna functions differ from those in Ref.~\cite{Gehrmann-DeRidder:2005svg} only in finite pieces.  

However, a residual dependence on the choice of single-real antennae is left in the construction of double-real antenna functions and the real-virtual antenna functions constructed in this paper. 
The deepest $1/\e^4$ and $1/\e^3$ poles correspond to the universal unresolved behaviour and are identical, but the $1/\e^2$ and $1/\e$ poles are potentially different.  This is understood as finite differences in the single-real antennae and should pose no issues in application to the antenna-subtraction scheme, when used with a consistent set of $X_3^0$, $X_4^0$, and $X_3^1$ antenna functions.
Indeed, this is the case, and we find that Eqs.~\eqref{eq:J22QQ}--\eqref{eq:J22htGG} satisfy Eq.~\eqref{eq:J22I2}, thereby demonstrating the consistency of the NNLO antenna subtraction scheme based on the antennae derived directly from the desired singular limits presented here and in Ref.~\cite{paper2}.

\section{Outlook}
\label{sec:outlook}

In this paper, we have extended the algorithm to build higher-order antenna functions presented in our previous publication Ref.~\cite{paper2} to the case where explicit poles in $\e$ are present, pertaining to mixed real and virtual corrections in higher-order calculations.
As a proof of the applicability of our new method, we have explicitly derived all $X_3^1$ antenna functions describing real-virtual radiation. Together with the real- and double-real antenna functions, $X_3^0$ and $X_4^0$, derived in our previous work Ref.~\cite{paper2}, this completes the derivation of a new consistent set of antenna functions for NNLO calculations of processes with massless partons in the final state.

We have identified two complementary sets of design principles relating to mixed real and virtual antenna functions which we dubbed generic and subtraction-scheme-dependent design principles. While the former ensures that each antenna function has judicious physical properties and obeys all unresolved limits, the latter matches the mixed real-virtual antennae onto the full antenna subtraction scheme. 
We have explicitly verified the consistency of the method at NNLO by recalculating the integrated dipoles $\J{2}$ in the new scheme. These have a direct relation to the general Catani IR singularity operators, and therefore the pole structure of one- and two-loop matrix elements.

The completion of a consistent set of improved antenna functions for (double-)real and real-virtual radiation provides an essential step towards the automation of the antenna-subtraction method at NNLO. This will also be essential for going beyond the current state-of-the-art and addressing complicated processes such as $e^+e^- \to 4$ jets.
We believe that, apart from reducing the complexity of subtraction terms, our new antenna functions will reduce the computational overhead associated with precision calculations. Our assessment is based on the fact that we have chosen our design principles in such a way that they avoid the need for spurious subtraction terms as much as possible.

In order to address precision phenomenology at hadron colliders such as the LHC, NNLO antenna functions for initial-final (IF) and initial-initial (II) configurations are required. These can be constructed using the same algorithm set-out here and in Ref.~\cite{paper2}, using the known IF and II unresolved limits.
For the cancellation of poles in virtual and double-virtual matrix elements, those antenna functions then also need to be integrated over the respective initial-final and initial-initial antenna phase spaces. We note that a first step in this direction has recently been taken and the NLO antennae for the IF and II configurations have been derived using this approach in Ref.~\cite{Fox:2023bma}. A full antenna subtraction scheme would streamline the calculation of full-colour $pp \to 3$~jets as well as bring other processes such as $pp \to V$+2 jets, $pp \to H$+2 jets into scope (as and when the two loop matrix elements become available).

We have also so far focused our work on antennae with massless particles. Calculations involving massive quarks, such as bottoms or tops, require antenna functions with massive particles. Again, these antennae need to be integrated over their respective massive antenna phase spaces so that they can be used to cancel explicit poles in virtual matrix elements.

We wish to stress that the algorithm presented in this and our previous paper Ref.~\cite{paper2} can straightforwardly be promoted to \NthreeLO calculations, provided that the appropriate unresolved limits are known analytically. The definition of an antenna-subtraction scheme for \NthreeLO calculations will require three new types of antenna functions, namely triple-real (RRR), double-real-virtual (RRV), and real-double-virtual (RVV). While the removal of overlapping singularities at \NthreeLO may be tedious in practice, we anticipate that our algorithm provides a suitable baseline for these endeavours.

\acknowledgments
We thank Xuan Chen, Aude Gehrmann-De Ridder, Thomas Gehrmann, Matteo Marcoli and Kay Schönwald for enlightening discussions and helpful advice. OBW and NG thank the University of Zurich, and especially Thomas Gehrmann and his research group for their kind hospitality, while visiting the University of Zurich. This visit was supported in part by the Pauli Center for Theoretical Studies, in part by the UK Science and Technology Facilities Council under contract ST/T001011/1 and in part by the Swiss National Science Foundation under contract 200021-197130.

\appendix

\section{Tree-level single unresolved limits}
\label{app:treelimits}

The tree-level soft factor is given by the eikonal factor for particle $b$ radiated between two hard radiators ($a$ and $c$),
\begin{align}
    \Sgzero(i^h,j,k^h) &=    \frac{2s_{ik}}{s_{ij}s_{jk}},  \\
    S_q^{(0)}(i^h,j,k^h) &= S_{\bar{q}}^{(0)}(i^h,j,k^h) =0.
\end{align}

The tree-level splitting functions $P_{ab}^{(0)}(i^h,j)$ are {\em not singular} in the limit where the hard radiator $a$ becomes soft and are related to the usual spin-averaged splitting functions, cf.~\cite{Altarelli:1977zs,Dokshitzer:1977sg}, by
\begin{align} 
\label{eqn:Pqg}
\Pqgzero(i^h,j) &= \frac{1}{s_{ij}} \Pqg(\xj), \\
\Pqgzero(i,j^h) &= 0,\\
\label{eqn:Pqq}
\Pqqzero(i^h,j) &= \frac{1}{s_{ij}} \Pqq(\xj),\\
\Pqqzero(i,j^h) &= \frac{1}{s_{ij}} \Pqq\omxj,\\
\label{eqn:Pgg}
\Pggzero(i^h,j) &= \frac{1}{s_{ij}} \PggS(\xj),\hfill\\
\Pggzero(i,j^h) &= \frac{1}{s_{ij}}\PggS\omxj ,
\end{align}
with
\begin{eqnarray}
\label{eq:Pqg}
\Pqg(\xj) &=& \left(\frac{2\omxj}{\xj} + \ome \xj \right) , \\
\label{eq:Pqq}
\Pqq(\xj) &=& \left( 1 -\frac{2\omxj\xj}{\ome} \right) = \Pqq\omxj , \\
\label{eq:PggS}
\PggS(\xj)&=& \left( \frac{2\omxj}{\xj} + \xj \omxj \right) , 
\end{eqnarray}
and
\begin{equation}
\PggS(\xj) + \PggS\omxj \equiv \Pgg(\xj).
\end{equation}
Here, the momentum fraction $\xj$ is defined with reference to the third particle in the antenna, $\xj = s_{jk}/(s_{ik}+s_{jk})$.

\section{Integration of $X_3^1$}
\label{app:integration}

The integrated antenna is obtained by integrating over the antenna phase space,
\begin{equation}
\label{eq:phijkFF}
{\cal X}_{3}^1(s_{ijk}) =
\left(8\pi^2\left(4\pi\right)^{-\e} e^{\e\gamma}\right)
\int {\rm d}\, \Phi_{X_{ijk}} X_{3}^1,
\end{equation}
with $d=4-2\e$.
As in Ref.~\cite{Gehrmann-DeRidder:2005btv}, we have included a normalisation factor to account for powers of the QCD coupling constant. 
The antenna phase space is given by
\begin{equation}
    {\rm d}\, \Phi_{X_{ijk}}  = \frac{1}{16\pi^2}\frac{1}{\Gamma(1-\e)} \left(\frac{4\pi}{s_{ijk}}\right)^{\e} s_{ijk} {\rm d} I ,
 \end{equation}
 with
 \begin{align}
{\rm d} I &= {\rm d}y_{ij}  {\rm d}y_{jk}
\left(y_{ij} y_{jk} (1-y_{ij}-y_{jk})\right)^{-\e} ,
\end{align}
where $0 < y_{ij} < 1$ and $0 < y_{jk} < 1-y_{ij}$.
Setting $y_{jk} = (1-y_{ij}) z$, then
\begin{align}
{\rm d}I &= {\rm d}y_{ij} \, {\rm d}z \, y_{ij}^{-\e} \, \left(1-y_{ij}\right)^{1-2\e} 
 \, z^{-\e} \, \omz^{-\e} ,
\end{align}
with $0 < y_{ij},z < 1$.

The integrals we encounter are of the form,
\begin{align}
&\int_0^1 {}_2F_1(\pm\e,\pm\e,1\pm\e,z)   z^{\alpha} \omz^{\beta}
{\rm d}z \\
& \hspace{2cm}= 
\frac{\Gamma(\alpha+1)\Gamma(\beta+1)}{\Gamma(\alpha+\beta+2)}{}_3F_2 (\pm\e,\pm\e,\alpha+1,1\pm\e,\alpha+\beta+2,1), \nonumber \\
&\int_0^1 {}_2F_1(\pm\e,\pm\e,1\pm\e,1-z)   z^{\alpha} \omz^{\beta}
{\rm d}z \\
& \hspace{2cm}= 
\frac{\Gamma(\alpha+1)\Gamma(\beta+1)}{\Gamma(\alpha+\beta+2)}{}_3F_2 (\pm\e,\pm\e,\beta+1,1\pm\e,\alpha+\beta+2,1), \nonumber \\
&\int_0^1 z^{\alpha} \omz^{\beta}
{\rm d}z = \frac{\Gamma(\alpha+1)\Gamma(\beta+1)}{\Gamma(\alpha+\beta+2)} .
\end{align}

\section{Integrals of $X_3^1$ antennae derived using the $X_3^0$ of Ref.~\cite{Gehrmann-DeRidder:2005btv}}
\label{app:X31oldX30}

\begingroup
\allowdisplaybreaks
In this appendix, we list the integrals over the antenna phase space of the renormalised $X_3^1$ antenna constructed using the $X_3^0$ antennae of Ref.~\cite{Gehrmann-DeRidder:2005btv}.

\begin{eqnarray}
\label{eq:A31intoldX3}
{\cal A}_3^1 (s_{ijk}) &=& S_{ijk}^2\Biggl [
-\frac{1}{4\e^4}
-\frac{31}{12\e^3}
+\frac{1}{\e^2} \left(
-\frac{53}{8}
+\frac{11}{24}\pi^2
\right)
+\frac{1}{\e} \left(
-\frac{647}{24}
+\frac{22}{9}\pi^2
+\frac{23}{3}\zeta_3
\right) \nonumber \\
&& \hspace{2cm}
 + \left(
-\frac{1289}{12}
+\frac{199}{24}\pi^2
+\frac{635}{18}\zeta_3
+\frac{13}{1440}\pi^4
\right)
 + \order{\e}\Biggr] \, ,\\
\label{eq:A31tintoldX3}
\widetilde{{\cal A}}_3^1 (s_{ijk}) &=& S_{ijk}^2\Biggl [
+\frac{1}{\e^2} \left(
-\frac{5}{8}
+\frac{1}{6}\pi^2
\right)
+\frac{1}{\e} \left(
-\frac{19}{4}
+\frac{1}{4}\pi^2
+7\zeta_3
\right)
\nonumber \\
&& \hspace{2cm}
 + \left(
-\frac{435}{16}
+\frac{29}{16}\pi^2
+\frac{21}{2}\zeta_3
+\frac{7}{60}\pi^4
\right)
 + \order{\e}\Biggr] \, ,
\\
\label{eq:A31hintoldX3}
\widehat{{\cal A}}_3^1 (s_{ijk}) &=& S_{ijk}^2\Biggl [
+\frac{1}{3\e^3}
+\frac{1}{2\e^2}
+\frac{1}{\e} \left(
\frac{19}{12}
-\frac{7}{36}\pi^2
\right)
+ \left(
\frac{109}{24}
-\frac{7}{24}\pi^2
-\frac{25}{9}\zeta_3
\right)
\nonumber \\
&& \hspace{2cm}
 + \order{\e}\Biggr] \, ,
\\
\label{eq:d31intoldX3}
{\cal D}_3^1 (s_{ijk}) &=& S_{ijk}^2\Biggl [
-\frac{1}{4\e^4}
-\frac{8}{3\e^3}
+\frac{1}{\e^2} \left(
-\frac{1193}{144}
+\frac{13}{24}\pi^2
\right)
+\frac{1}{\e} \left(
-\frac{8473}{216}
+\frac{49}{18}\pi^2
+\frac{73}{6}\zeta_3
\right)
\nonumber \\
&& \hspace{2cm}
 + \left(
-\frac{18937}{108}
+\frac{9485}{864}\pi^2
+\frac{535}{12}\zeta_3
+\frac{79}{480}\pi^4
\right)
 + \order{\e}\Biggr] \, ,
\\
\label{eq:d31tintoldX3}
\widetilde{{\cal D}}_3^1 (s_{ijk}) &=& S_{ijk}^2\Biggl [
+\frac{1}{\e^2} \left(
-\frac{5}{16}
+\frac{1}{12}\pi^2
\right)
+\frac{1}{\e} \left(
-\frac{13}{6}
+\frac{11}{72}\pi^2
+\frac{5}{2}\zeta_3
\right)\nonumber \\
&& \hspace{2cm}
 + \left(
-\frac{395}{36}
+\frac{941}{864}\pi^2
+\frac{55}{12}\zeta_3
-\frac{7}{180}\pi^4
\right)
 + \order{\e}\Biggr] \, ,
\\
\label{eq:d31hintoldX3}
\widehat{{\cal D}}_3^1 (s_{ijk}) &=& S_{ijk}^2\Biggl [
+\frac{1}{3\e^3}
+\frac{5}{9\e^2}
+\frac{1}{\e} \left(
\frac{139}{72}
-\frac{7}{36}\pi^2
\right)
+ \left(
\frac{443}{72}
-\frac{35}{108}\pi^2
-\frac{25}{9}\zeta_3
\right)\nonumber \\
&& \hspace{2cm}
 + \order{\e}\Biggr] \, ,
\\
\label{eq:E31intoldX3}
{\cal E}_3^1 (s_{ijk}) &=& S_{ijk}^2\Biggl [
+\frac{11}{18\e^2}
+\frac{1}{\e} \left(
\frac{74}{27}
-\frac{1}{9}\pi^2
\right)
+ \left(
\frac{1441}{108}
-\frac{149}{216}\pi^2
-4\zeta_3
\right)
 + \order{\e}\Biggr] \, ,
\\
\label{eq:E31tintoldX3}
\widetilde{{\cal E}}_3^1 (s_{ijk}) &=& S_{ijk}^2\Biggl [
+\frac{1}{6\e^3}
+\frac{35}{36\e^2}
+\frac{1}{\e} \left(
\frac{509}{108}
-\frac{1}{4}\pi^2
\right)
+ \left(
\frac{1670}{81}
-\frac{35}{24}\pi^2
-\frac{31}{9}\zeta_3
\right)\nonumber \\
&& \hspace{2cm}
 + \order{\e}\Biggr] \, ,
\\
\label{eq:E31hintoldX3}
\widehat{{\cal E}}_3^1 (s_{ijk}) &=& S_{ijk}^2\Biggl [
+\frac{1}{3\e}
+ \left(
\frac{172}{81}
-\frac{11}{108}\pi^2
\right)
 + \order{\e}\Biggr] \, ,
\\
\label{eq:f31intoldX3}
{\cal F}_3^1 (s_{ijk}) &=& S_{ijk}^2\Biggl [
-\frac{1}{4\e^4}
-\frac{11}{4\e^3}
+\frac{1}{\e^2} \left(
-\frac{85}{9}
+\frac{5}{8}\pi^2
\right)
+\frac{1}{\e} \left(
-\frac{9827}{216}
+\frac{55}{18}\pi^2
+\frac{44}{3}\zeta_3
\right)
\nonumber \\
&& \hspace{2cm}
 + \left(
-\frac{88961}{432}
+\frac{5665}{432}\pi^2
+\frac{473}{9}\zeta_3
+\frac{181}{1440}\pi^4
\right)
 + \order{\e}\Biggr] \, ,
\\
\label{eq:f31hintoldX3}
\widehat{{\cal F}}_3^1 (s_{ijk}) &=& S_{ijk}^2\Biggl [
+\frac{1}{3\e^3}
+\frac{11}{18\e^2}
+\frac{1}{\e} \left(
\frac{19}{9}
-\frac{7}{36}\pi^2
\right)
+ \left(
\frac{167}{24}
-\frac{77}{216}\pi^2
-\frac{25}{9}\zeta_3
\right)\nonumber \\
&& \hspace{2cm}
 + \order{\e}\Biggr] \, ,
\\
\label{eq:G31intoldX3}
{\cal G}_3^1 (s_{ijk}) &=& S_{ijk}^2\Biggl [
+\frac{11}{18\e^2}
+\frac{1}{\e} \left(
\frac{169}{54}
-\frac{1}{9}\pi^2
\right)
+ \left(
\frac{3355}{216}
-\frac{161}{216}\pi^2
-4\zeta_3
\right)
 + \order{\e}\Biggr] \, ,
\\
\label{eq:G31tintoldX3}
\widetilde{{\cal G}}_3^1 (s_{ijk}) &=& S_{ijk}^2\Biggl [
+\frac{1}{6\e^3}
+\frac{41}{36\e^2}
+\frac{1}{\e} \left(
\frac{325}{54}
-\frac{1}{4}\pi^2
\right)
+ \left(
\frac{9053}{324}
-\frac{41}{24}\pi^2
-\frac{31}{9}\zeta_3
\right)\nonumber \\
&& \hspace{2cm}
 + \order{\e}\Biggr] \, ,
\\
\label{eq:G31hintoldX3}
\widehat{{\cal G}}_3^1 (s_{ijk}) &=& S_{ijk}^2\Biggl [
+\frac{7}{18\e}
+ \left(
\frac{895}{324}
-\frac{11}{108}\pi^2
\right)
 + \order{\e}\Biggr] \, .
\end{eqnarray}

\endgroup
For the $A$-type, $E$-type and $G$-type antennae, we find complete agreement with the pole structure of the analogous integrated antennae given in Ref.~\cite{Gehrmann-DeRidder:2005btv}.  For the $D$-type and $F$-type antennae, we have utilised the $X_3^0$ sub-antenna given in Eqs.~(6.13) and (7.13) of Ref.~\cite{Gehrmann-DeRidder:2005btv} respectively and therefore the pole structures of the combinations $2 \left( {{\cal D}}_3^1+\widetilde{{\cal D}}_3^1 \right)$,$2 \widehat{{\cal D}}_3^1$, $3 {{\cal F}}_3^1$ and $3 \widehat{{\cal F}}_3^1$ agrees with the expressions for $\Drvold$, $\Dhrvold$, $\Frvold$ and $\Fhrvold$ respectively,
given by Eqs. (6.22), (6.23), (7.22) and (7.23) of Ref.~\cite{Gehrmann-DeRidder:2005btv} respectively to $\order{\e^0}$.

\bibliographystyle{jhep}
\bibliography{bib2}{}
\end{document}